\documentclass[prd,aps,twocolumn,a4paper,showkeys,nofootinbib,floatfix]{revtex4-1}

\usepackage{graphicx,psfrag}
\usepackage{mathrsfs}
\usepackage{amsmath,amsfonts,amssymb}
\usepackage{multirow}
\usepackage{comment}
\usepackage{ulem}
\usepackage{hyperref}
\usepackage{enumitem}
\usepackage{morefloats}
\usepackage{bm}

\newcommand{\be}{\begin{equation}}
\newcommand{\ee}{\end{equation}}
\newcommand{\bea}{\begin{eqnarray}}
\newcommand{\eea}{\end{eqnarray}}
\newcommand{\bel}{\begin{align}}
\newcommand{\eel}{\end{align}}

\def\dee{{\rm d}}

\def\Msun{{\rm M_{\odot}}}
\def\GMc2{{\rm G M_{\odot} c^{-2}}}

\def\M{\mathcal{M}}

\def\Iloss{L_{\tilde\Lambda\tilde\Lambda}}

\def\Mo{{\rm M_{\odot}}}

\def\TEOB{{\tt TEOBResumS}}
\def\TF2{{\tt TaylorF2}}

\def\NRT1{{\tt IMRPhenomPv2NRTidal}}

\def\params{\boldsymbol{\theta}}
\def\tLam{\tilde{\Lambda}}
\def\tLamB{\tilde{\Lambda}_E}

\usepackage{pifont}
\newcommand{\cmark}{\ding{51}}%
\newcommand{\xmark}{\ding{55}}%

\usepackage{color}
\definecolor{cyan}{rgb}{0,0.9,0.9}
\definecolor{orange}{rgb}{0.9,0.5,0}
\definecolor{magenta}{rgb}{1,0,1}
\definecolor{purple}{rgb}{0.8,0.4,0.8}
\definecolor{gray}{rgb}{0.8242,0.8242,0.8242}

\begin{document}

\title{Waveform systematics in the gravitational-wave inference of\\ tidal parameters and equation of state from binary neutron star signals}

\author{Rossella \surname{Gamba}$^{1}$}
\author{Matteo \surname{Breschi}$^{1}$}
\author{Sebastiano \surname{Bernuzzi}$^{1}$}
\author{Michalis \surname{Agathos}$^{1}$}
\author{Alessandro \surname{Nagar}$^{2}$}

\affiliation{${}^1$Theoretisch-Physikalisches Institut, Friedrich-Schiller-Universit{\"a}t Jena, 07743, Jena, Germany}
\affiliation{${}^2$ INFN Sezione di Torino, Via P. Giuria 1, 10125 Torino, Italy}
\affiliation{${}^{3}$ Dipartimento di Fisica, Universit\`a di Torino, via P. Giuria 1, 10125 Torino, Italy}
\date{\today}

\begin{abstract}
  
  Gravitational-wave signals from binary neutron star coalescences
  carry information about the star's equation of state in their tidal signatures.
  A major issue in the inference of the tidal parameters (or directly of the equation
  of state) is the systematic error introduced by the waveform approximants. 
  We use a bottom-up approach based on gauge-invariant phase analysis
  and the Fisher information matrix to investigate waveform systematics
  and help identifying biases in parameter estimation.
  A mock analysis of 15 different binaries indicates that systematics
  in current waveform models dominate over   
  statistical errors at signal-to-noise ratio (SNR) ${\gtrsim} 80$.
  This implies biases in the inference of the reduced tidal parameter
  that are are larger than the
  statistical $90\%$ credible-intervals. 
  For example, while the neutron-star radius could be constrained at
  ${\sim} 5\%$ level at SNR 80, systematics can be at 
  the ${\sim} 10\%$ level. 
  We apply our approach to GW170817 (SNR ${\sim}30$) and confirm  
  that no significant systematic effects are present. 
  Using an optimal frequency range for the analysis, 
  we estimate a neutron-star radius of $12.5^{+1.1}_{-1.8}\,$km. The
  latter is consistent with an electromagnetic-informed prior and the recent
  NICER measurement.  
  Exploring SNR ${\gtrsim}100$ in view of third-generation detectors, we find
  that all the current waveform models lead to differences of at
  least 1-sigma in the inference of the reduced tidal parameter (for any
  value of the latter).  
  We conclude that current waveform models, including those from
  numerical relativity, are insufficient to infer the equation of
  state in the loudest (and potentially most informative)
  events that will be observed by advanced and third generation
  detectors.     
\end{abstract}


\maketitle

\section{Introduction}
\label{sec:intro}

The detection of GW170817~\cite{TheLIGOScientific:2017qsa}, 
the first coalescing binary neutron star (BNS) system seen by LIGO-Virgo detectors, 
demonstrated how gravitational waves (GWs) can be employed as a mean to investigate the
properties of cold, dense matter~\cite{Radice:2017lry, Abbott:2018exr, Essick:2019ldf, LIGOScientific:2019eut}. 
Parameter estimation (PE) of GW data gives direct
information on the masses, spins and tidal parameters of the two objects involved in the coalescence. 
Matched filtering analyses are performed in the Fourier domain by matching
the whitened data, i.e time series sampled at a constant sampling frequency, to a large number of template waveforms 
within a Bayesian framework~\cite{Ghosh:2015jra}. 
The tools employed during PE are based 
on Markov-chain Monte Carlo (MCMC) methods or nested sampling
algorithms~\cite{Veitch:2014wba}.
The template waveforms are obtained relying on approximate solutions 
of the two body problem in general relativity 
(see e.g. Refs.~\cite{Flanagan:2007ix,Damour:2009wj,Damour:2011xga,Damour:2015isa}
and references therein). Different approximations and methods give  
rise to different template families, which - during the process of PE - may in principle 
lead to different results in the recovery of the source parameters. 
Errors and biases completely referable to waveform modeling choices 
are commonly labelled as {\it waveform systematics}, and are the main topic of the present paper.

Significant waveform systematics, larger than statistical uncertainties,
have yet to be observed for binary neutron star systems: 
looking at the results coming from the recent observations of BNS mergers,
the parameters of both GW190425~\cite{Abbott:2020uma} and GW170817~\cite{LIGOScientific:2018mvr} have been 
demonstrated to be largely consistent between different 
waveform families. 
However, recent studies~\cite{Dudi:2018jzn, Samajdar:2018dcx, Samajdar:2019ulq, Messina:2019uby, Agathos:2019sah, Narikawa:2019xng, Chen:2020fzm} 
have pointed out that the measured tidal parameters can be strongly biased 
depending on the employed tidal and point mass descriptions of the waveform approximant.
The agreement between the different GW models employed
in the PE of the observed BNS signals
is then mainly due to the relatively low signal-to-noise 
ratio (SNR). With the increasing sensitivities of next-generation 
detectors~\cite{Sathyaprakash:2011bh,TheLIGOScientific:2014jea,Aasi:2013wya,Harry:2010zz,TheVirgo:2014hva,Maggiore:2019uih}, 
waveform systematics will affect the measurements thus leading to discordant (or inconclusive) results.

In this context, the necessity of understanding the errors introduced by 
waveform systematics arises. In this paper, we aim at tackling
the issue for SNRs relevant for advanced and 3G detectors,
and provide a bottom-up approach to guide future BNS analyses.
In particular, in Sec.~\ref{sec:ana} we summarize the current knowledge 
of the theoretical tools which are employed to a-priori predict the 
presence of waveform systematics, we expand on the argument of 
Ref.~\cite{Messina:2019uby} and propose a way, inspired 
by~\cite{Cutler:2007mi, Favata:2013rwa}, to estimate the bias 
that may affect tidal parameters. 
In Sec.~\ref{sec:wf} we summarize the key features of the GW models used in our analysis,
and compare them by computing their gauge-invariant phasing.
In Sec.~\ref{sec:analysis} we perform mock PE experiments
({\it injections}) with 15 binaries having signal-to-noise ratio ${\sim}80$, 
to study the posterior distributions of the typical parameters of 
interest of a BNS merger, such as tidal deformabilities, mass ratio and spins.
Differently from previous studies, we focus on injections of
different masses and EOS (Cf. \cite{Dudi:2018jzn, Samajdar:2018dcx,
  Samajdar:2019ulq} where fewer binaries have been considered) and
nonspinning waveforms  (See \cite{Samajdar:2019ulq} for spin effects.)
In particular, we discuss the impact of waveform systematics on the
inference of the tidal parameters and the estimation of the radii of the single NSs.
In Sec~\ref{sec:gw170817} we apply the methods developed during the previous
sections to GW170817. We re-analyze the event's data and find that 
analyses considering up to 1kHz are free of systematics.
Finally, in Sec.~\ref{sec:3G} we estimate the impact of waveform systematics
for BNS events detected with third generation detectors and find that
statistical errors will be comparable to waveform systematics from SNRs
$>200$ for $\tLam \simeq 400 - 1000$.

Throughout the whole paper we label the two bodies as $A, B$. 
We denote the component masses as $m_A, m_B$, the dimensionless spins of the bodies as $\chi_{A, B}$,
the total mass as $M = m_A + m_B$, and define the chirp mass of the binary as $\mathcal{M}_c = (m_A m_B)^{3/5}/(M)^{1/5}$.
We define the quadrupolar tidal parameters as
\be
\Lambda_A \equiv \frac{2}{3}\, \mathcal{C}_A^{-5} k^{(2)}_A \label{eq:Lambda_A},
\ee
where $k_A^{(2)}$ is the dimensionless gravitoelectric Love number \cite{Hinderer:2007mb, Damour:2009vw}, and $\mathcal{C}_A \equiv Gm_A/(c^2R_A)$ is the compactness parameter.
$\Lambda_A$ is also denoted by $\bar\lambda_2$ \cite{Yagi:2013sva}.
The quadrupole tidal parameters enters at the leading order in the phase of the waveform through the reduced tidal parameter \cite{Flanagan:2007ix,Favata:2013rwa}
\be
\tilde\Lambda = \frac{16}{13}\frac{(m_A+12 m_B)m_A^4 \Lambda_A}{M^5}+(A\leftrightarrow B)\,.
\label{eq:Lambda_tilde}
\ee
We often switch between mass-rescaled
quantities in geometrical units $c=G=1$ and physical units. Since $GM_\odot\simeq4.925490947\,\mu$s or
${\simeq}1.476625038\,$km, the dimensionless frequency $\hat{\omega}=GM\omega$
relates to the frequency in Hz by 
\be
f = \frac{\omega}{2\pi}\simeq32.3125\, \hat{\omega}\,\frac{M_\odot}{M}\,\mbox{kHz}\,.
\ee

\section{Origin of systematics}
\label{sec:ana}

Waveform systematics are intrinsically related to the concept of
measurability of the waveform parameters. They arise
when the differences due to template choice are larger than those
induced by noise fluctuations in the detector and statistical uncertainties, 
i.e when the distributions of the estimated parameters $\bar{\params}$ have a width
$\sigma_{\params}$ smaller than the differences $\Delta \params$
induced by waveform models. 
In this section we highlight, with basic analytical arguments, that the
systematics on tidal parameters crucially depend on the frequency
regime at which the measurement is effectively performed.

Optimal gravitational-wave data analysis of compact binaries are based
on matched-filtering techniques in which the data are ``best matched''
to waveform templates \cite{Owen:1998dk}. The accuracy requirements on the
waveforms used in the matched filtering depend on whether waveform
models are employed for detection or parameter
estimation. In the former
case, waveforms are required to be only effectual while 
in the latter they are required to be faithful~\cite{Damour:1997ub}.
To quantify these concepts, it is necessary to introduce a metric in the waveform
space in order to measure how close two waveforms are.
The basic quantity used in in GW analysis theory is 
the Wiener inner product between two
waveforms $h(t)$ and $k(t)$, defined by
\be
\label{eq:innerprod}
(h|k) = 4 \Re \int \frac{\tilde{h}(f) \, \tilde{k}^*(f)}{S_n(f)}\,{\rm d}f\, ,
\ee
where $\tilde{h}(f)$ is the Fourier transform of $h(t)$ and
$S_n(f)$ is the power spectral density (PSD) of the detector.
The faithfulness (or match) is the normalized and noise-weighted inner
product
\be
\label{eq:faithfulness}
\mathcal{F} = \max_{t_c, \phi_c} \frac{(h|k)}{\sqrt{(h|h) (k|k)}}\,,
\ee
where $t_c, \phi_c$ are respectively the time and phase 
of the waveform at a reference time.
The match $\mathcal{F}$ defines an ``angle'' 
in the waveform space; $\mathcal{F}=1$ indicates perfect overlap
between $h$ and $k$.
The mismatch $\bar{\mathcal{F}}=1-\mathcal{F}$ gives the loss in
signal-to-noise ratio (squared) when the waveforms are aligned in time
and phase. 
Accuracy requirements for both detection and PE can be expressed in
terms of $\mathcal{F}$~\footnote{Although this has become a common practice,
  it would be more appropriate to express these requirements by means
  of suitable effectualeness, faithfulness, and accuracy functional,
  see~\cite{Damour:2010zb}.}. 
A mismatch of $\bar{\mathcal{F}}=0.03$ corresponds to ${\sim}10$\% of detection
losses \cite{Lindblom:2008cm}, which is assumed as the effectualness
condition for a template bank.
Necessary conditions for faithful waveform models can be expressed in terms of $\mathcal{F}$
\cite{Lindblom:2008cm,Damour:2010zb} (see below).

Generally, the parameters of a GW signal are measured 
using matched-filtering techniques within a Bayesian framework~\cite{Veitch:2014wba}.
Defining $\tilde{d}(f) = \tilde{A}_d(f) e^{i \Psi_d(f)}$ as the target (injected or measured) strain of data, 
$\tilde{h}(f,\params) = \tilde{A}_h(f,\params) e^{i \Psi_h(f,\params)}$ as the template waveform and $\params$ as the set of parameters
on which $h$ depends, the likelihood function is
\be
\label{eq:likelihood}
p(d|\params) \propto e^{-\frac{1}{2}(d-h|d-h)} \,.
\ee
Writing $(d-h|d-h)=(d|d) + (h|h) -2(d|h)$, 
the maximization of the likelihood
can be interpreted as the maximization of the matched-filtered SNR, 
\be\label{eq:rho2}
\rho =  \frac{(d|h)}{\sqrt{(h|h)}} \propto \Re \int \frac{\tilde{A}_d \tilde{A}_h e^{i \Delta \Psi}}{S_n} {\rm d}f\,,
\ee
where $\Delta\Psi = \Psi_h - \Psi_d$.
The SNR $\rho$ quantifies the amount of signal deposited in the recorded
data $d$ that is matching a given template $h$.
The \textit{optimal} SNR, instead, is defined as 
the matched-filtered SNR computed within the assumption $d\approx h$, 
\be\label{eq:rhoopt}
\rho_{\rm opt} =  \sqrt{(h|h)}\,.
\ee
This value identifies the SNR 
we would get if 
the signal were coincident with the template
and the noise realization were identically zero, 
which is a first order approximation of the actual matched-filtered value.
Indeed, assuming $d=h+n$ and expanding around small values of $n$, 
we get $\rho = \rho_{\rm opt} + O(n)$. 
GW data analysis delivers probability distributions of the sampled parameters
(posteriors), which can be characterized by their maximum probability (peak) values
and credible intervals. The measurements thus obtained can be affected by 
statistical uncertainties due to fluctuations of the detector 
noise, which impact the posteriors by widening the credible intervals, and 
systematic effects due to the waveform models employed, which
can can influence PE by shifting the posterior distributions
with respect to the true values.

\subsection{Statistical Errors}
\label{sec:staterr}

Under the assumption of Gaussian noise, the variance
$\sigma^2_{\theta_i}$ on the measurement of a generic parameter
$\theta_i$ due to statistical errors can be computed
through the Fisher information matrix $F_{ij}$  (see e.g. \cite{Cutler:1994ys,Lindblom:2008cm,Damour:2010zb}).
Given a waveform model $\tilde{h}(f, \theta) = \tilde A_h(f) e^{i \Psi_h(f)}$, the element $(i,j)$ of $F$ is defined as
\be
\label{eq:Fij}
F_{ij} = (\partial_i h | \partial_j h)\simeq 4 \int \frac{\tilde A_h^2}{S_n} \left(\partial_i \Psi_h \partial_j \Psi_h \right) {\rm d}f,
\ee
where $\partial_i = \frac{\partial}{\partial \theta_i}$ and in the last
equation we assume that the amplitude $\tilde{A}$ is not correlated to 
other parameters \cite{Cutler:1994ys, Poisson:1995ef}.
The variance of the distribution of $\theta_i$ can then be estimated from Eq.~\eqref{eq:Fij} as
\be
\label{eq:variance}
\sigma^{2}_{\theta_i} = (F^{-1})_{ii} \,.
\ee 

The Fisher Matrix formalism further allows one to identify the relevant frequency ranges at which different parameters are measured.
Focusing on Eq.~\eqref{eq:Fij}, it is clear that
the frequency ranges that contribute to the computation of $\sigma_{\theta_{i}}$
are those where the integrand
\be
\label{eq:integrand}
I_{ii}(f) = 4 \frac{\tilde{A}_h^2(f)}{S_n(f)}\,  \Big[ \partial_i \Psi_h(f)\, \partial_i \Psi_h(f) \Big]
\ee
is largest and different from zero. 
Using post-Newtonian (PN) waveforms, whose amplitude 
$\tilde{A}$ behaves as ${\sim} f^{-7/6}$, it is
immediate to show that on a logarithmic frequency axis these integrands are of
type $I_{ii} \sim f^{-4/ 3} S_{n}^{-1}(f) f^{p_i/3}$
where the exponent $p_i$ depends on the
particular parameter considered~\cite{Damour:2012yf}. 
For example, the chirp mass has $p_{\mathcal{M}}=-10$ and, for a fiducial equal-mass
$1.4+1.4\Mo$ BNS, is entirely determined by the signal at
low-frequencies ${\lesssim}30$~Hz. The symmetric mass ratio integrand
has $p_\nu=-6$ and the SNR integrand has $p_{\rm SNR}=0$, which implies that they are given by the
useful GW cycles below 50 and 100 Hz respectively for the fiducial
BNS (see e.g. Fig.~(3) of \cite{Damour:2012yf} and Fig.~(2) of \cite{Harry:2018hke}).
By contrast, the reduced tidal parameters has $p_{\tLam}=+10$,
i.e. $I_{\tLam\tLam}\sim f^2/S_n(f)$, that
indicates that the information on tides increases as ${\sim}f^2$
between 50 Hz and 800 Hz to then reach a finite limit at higher 
frequencies and decay after merger.
This marked difference in the frequency support of $I_{\tLam\tLam}$ indicates
that the measurement of $\tilde\Lambda$ is not
strongly correlated to that of the chirp mass and mass ratio~\cite{Damour:2012yf}, 
and that the magnitude of such a correlation decreases as signals become stronger, because the
tidal contribution becomes easier to distinguish from the rest of the signal  \cite{Damour:2012yf}.
Nonetheless, nontidal parameters can still impact the determination of $\tLam$ (see App.~\ref{app:PMTides}): the maximum likelihood values of $\tLam$ minimize the \textit{overall} high-frequency phase 
differences $\Delta \Psi$,
which can receive a non-negligible contribution from the point mass sectors of the approximants.

\begin{figure*}[t]
	\centering 
	\includegraphics[width=0.49\textwidth]{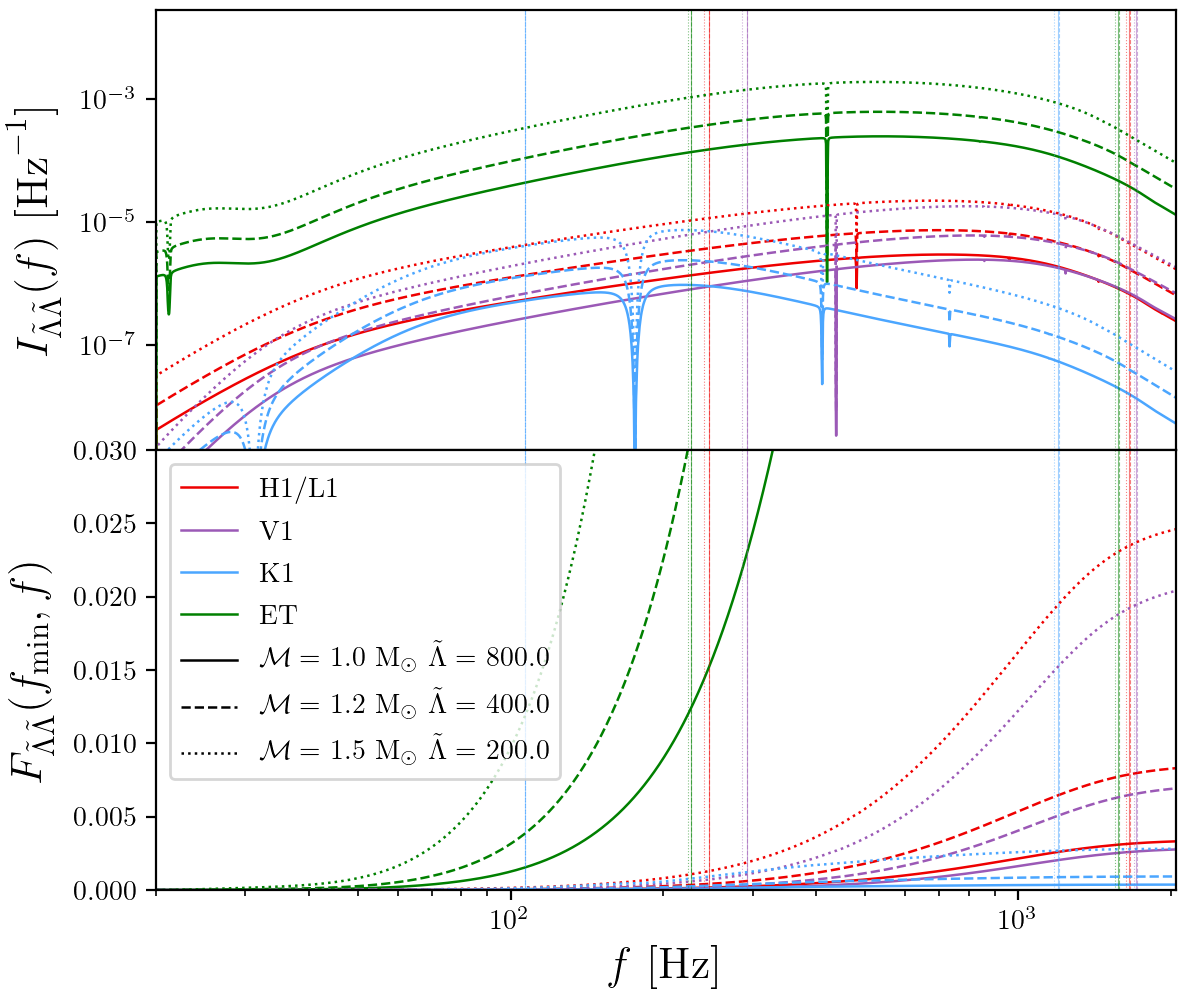}
	\includegraphics[width=0.49\textwidth]{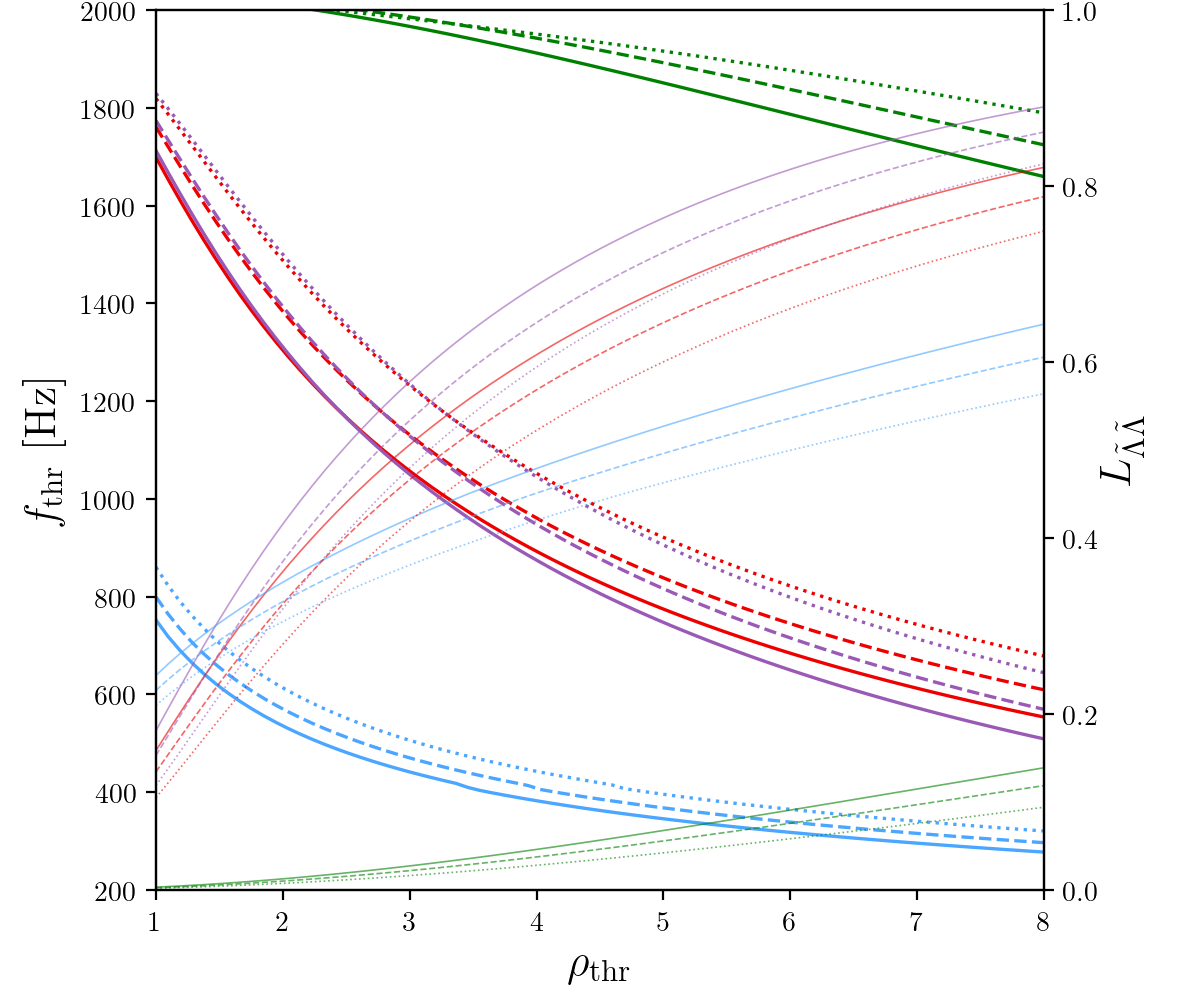}
	\caption{(Left panel) The figure shows $I_{\tLam\tLam}$ (top) and 
	  $F_{\tLam\tLam}(f_{\rm min}{=}\,20~{\rm Hz},f)$  (bottom)
	  computed for different combination of $(\mathcal{M},\tLam)$ 
	  fixing $|s_1|=|s_2|=0$, $q=1$, $D_L=40~{\rm Mpc}$, $\iota=0$
	  and locating the source in the optimal sky location for the involved detector
	  at the GPS time of GW170817 (1187008882.4). We employ ${\tt TaylorF2}$ to compute
	  the derivatives of the phase with respect to $\tLam$ and {\tt TEOBResumS} 
	  to account for corrections in the waveform amplitude.
	  The curves are estimated using design PSD expected for next-generation detectors:
	  red lines refer to LIGO design sensitivity~\cite{TheLIGOScientific:2014jea},
	  purple lines refer to Virgo design sensitivity~\cite{TheVirgo:2014hva}, 
	  blue lines refer to KAGRA design sensitivity~\cite{Aso:2013eba,Akutsu:2018axf},
	  and green lines refer to Einsten Telescope (configuration D) sensitivity~\cite{Punturo:2010zz,Hild:2010id}.
	  The vertical lines represent the frequencies $f_{5\%}^{\tLam}$, $f_{95\%}^{\tLam}$, 
	  defined in Eq.~\eqref{eq:fLamMinMax}.
	  (Right panel) Estimations of $f_{\rm thr}$ (thick lines) and $\Iloss$ (thin lines) as functions of $\rho_{\rm thr}\in[1,8]$
	  for the cases discussed in the left panel.}
        \label{fig:fM_inj}
\end{figure*}

\begin{comment}
Through the cumulative 
information $F_{ii}(f_{\rm min},f)$
it is then possible to find an optimal frequency range $[f^i_-,f^i_+]$
that encloses the relevant support of the integrand $I_{ii}(f)$ associated to $\theta_i$.
Indeed, the frequencies defined by 
\be
\label{eq:fLamMinMaxDiseq}
\frac{d}{df}\,F_{ii}(f_{\rm min},f) = I_{ii}(f)  > \frac{F_{ii}(f_{\rm min},f_{\rm max})}{f_{\rm max} - f_{\rm min}}
\ee
enclose the regions where the information gain is larger than the 
one we would have in case of uniformly distributed information.
We observe that the quantity
\be
\label{eq:meaninfo}
\bar I _{ii} = \frac{F_{ii}(f_{\rm min},f_{\rm max})}{f_{\rm max} - f_{\rm min}}\,,
\ee
is the mean value of the information distribution $I_{ii}$ estimated over the entire domain.
For inspiralling BNS mergers recorded with ground-based interferometers, 
$I_{ii}$ is expected to have vanishing tail contributions,
since the quantum and thermal noise contributions are dominant at low frequencies, 
while the signal amplitude has its maximum non-zero value at merger $f_{\rm mrg}\lesssim 2$~kHz~\cite{Breschi:2019srl}.
Subsequently, we can define the frequencies $f^i_\pm$
as the outermost values that satisfy the relation
\be
\label{eq:fLamMinMax}
I_{ii}(f=f^{i}_\pm)= \bar I _{ii}\,.
\ee
\end{comment}

The integrands $I_{ii}(f)$ can be employed to quantify the amount of information 
gathered on the parameter $\theta_i$ per frequency bin. We define the cumulative 
information gathered in an interval $[f_a,f_b]$ of the frequency domain as
\be
\label{eq:Icumulative}
F_{ii}(f_a,f_b)=\int_{f_a}^{f_b}  I_{ii}(f) \,{\rm d}f \,.
\ee
Values of $F_{ii}(f_a,f_b)$ close to zero indicate that the range $[f_a,f_b]$
does not include relevant information on $\theta_i$.
From Eq.~\eqref{eq:variance} we obtain
\be
\sigma_{\theta_i}^2\ge \frac{1}{F_{ii}(f_{\rm min}, f_{\rm max})}\,,
\ee
where $f_{\rm max}$ and $f_{\rm min}$ denote the upper and lower bounds of the frequency interval chosen for the analysis.

Through the information distribution $I_{ii}(f)$ and its integral $F_{ii}(f_{\rm min},f)$
it is possible to find an optimal frequency range 

where most of the information on $\theta_i$ is contained.
We define the upper frequency $f^i_{X\%}$ that encloses the $X\%$ of information 
on the $i$-th parameter from 
\be
\label{eq:fLamMinMax}
F_{ii}(f_{\rm min}, f^i_{X\%})= \frac{X}{100} \cdot F_{ii}(f_{\rm min}, f_{\rm max})\,.
\ee
This definition corresponds to the frequency of the $X^{\rm th}$ percentile of the information distribution $I_{ii}(f)$.
It is then possible to estimate the optimal frequency range for the measurement of the $i$-th parameter 
as the interval that encloses the 90\% of the total information, 
identified by the 5\% and the 95\% percentiles $[f_{5\%}^i,f_{95\%}^i]$.
Focusing on the tidal parameter $\tLam$,
Fig.~\ref{fig:fM_inj} (left panel) shows
the information distribution $I_{\tLam\tLam}(f)$ and the cumulative information $F_{\tLam\tLam}(f_{\rm min}, f)$
computed for some exemplary binary configurations 
using the expected design sensitivity curves for current ground-based detectors and $f_{\rm min}=20\,$Hz.
For fiducial BNS mergers with LIGO-Virgo design sensitivities, 
the optimal interval $[f_{5\%}^{\tLam},f_{95\%}^{\tLam}]$
spans a relatively high frequency range:
$f_{5\%}^{\tLam}\approx 300~{\rm Hz}$ 
and $f_{95\%}^{\tLam} > 1~{\rm kHz}$.
Note that the above intervals are independent of the distance of the source from
the detectors, i.e the same optimal frequency interval
pertains to a \textit{family} of signals with varying strengths and SNRs.

The key role played by distance and SNR does not lie in the determination of the
optimal interval, but rather in governing the extent to which the signal can be measured
and the parameters extracted.
Indeed, the accuracy on the $\tLam$ estimation depends crucially on the maximum frequency
at which we are able to discriminate the signal from noise fluctuations.
Within the assumption of $d\approx h$, it is possible to estimate 
the high-frequency threshold $f_{\rm thr}$ at which the signal power exceeds the noise contributions as
\be 
\label{eq:fM}
4\int_{f_{\rm thr}}^{f_{\rm  max}}  \frac{|\tilde{h}(f)|^2}{S_n(f)} {\rm d}f=  \rho^2_{\rm thr}\,,
\ee 
where $\rho_{\rm thr}$ is an arbitrary threshold value for the SNR.
In the case of multiple detectors, the integral in the left-hand side of Eq.~\eqref{eq:fM}
has to be replaced with the summation of the integrals evaluated on the different detectors,
as it is for an usual summation of SNRs.
With this definition, we guarantee that the power enclosed in the frequency range $[f_{\rm thr},f_{\rm max}]$ 
does not exceed the threshold $\rho_{\rm thr}$. 
The choice of $\rho_{\rm thr}$ is a subtle issue, 
since this value has to quantify the amount of power due to high-frequency noise and statistical fluctuations:
an overestimate will lead to the inclusion of portion of the signal as the noise contribution,
while, underestimating the threshold, 
one may be led to believe that the signal is more informative than it really is.

In principle, $\rho_{\rm thr}$ should be a negligible value compared   
to the total SNR and no signal should be gathered the range $[f_{\rm thr},f_{\rm max}]$.
A conservative choice is $\rho_{\rm thr} = 1$,
since it defines the range in which signal and noise contributions are comparable
and it ensures that we are not discarding a considerable amount of signal.
The choice of $\rho_{\rm thr}$ can be relaxed using the standard deviation 
estimated from the posterior distribution of the SNR coming from a PE analysis,
since it quantifies the uncertainty on the signal power.
Eq.~\eqref{eq:fM} is computed using the approximation of optimal SNR,
then the definition of $f_{\rm thr}$ is exact in the limit $d \to h$. 
In a realistic scenario, the noise contamination is non negligible,
and consequently $f_{\rm thr}$ can be interpreted as an upper frequency-bound
beyond which the signal power cannot exceed $\rho_{\rm thr}$;
this means that, even in the best case scenario $d\approx h$, the power 
enclosed in the frequency range $[f_{\rm thr}, f_{\rm max}]$ will always be lower 
(or equal, for $d=h$) than the threshold power defined by $\rho_{\rm thr}$.
Once $f_{\rm thr}$ is known, it is possible to evaluate the ratio 
\be 
\label{eq:Iloss}
L_{ii}(f_{\rm thr}) =\frac{ F_{ii}(f_{\rm thr},f_{\rm max})}{ F_{ii}(f_{\rm min},f_{\rm max})}\,.
\ee 
that quantifies the fractional information loss on the $i$-th parameter,
since $f_{\rm thr}$ represents, by construction, the maximum frequency at which the signal is relevant.
Fig.~\ref{fig:fM_inj} (right panel) shows the estimation of $f_{\rm thr}$ and $\Iloss$ as functions of $\rho_{\rm thr}$. As $\rho_{\rm thr}$ grows, 
$f_{\rm thr}(\rho_{\rm thr})$ decreases because a larger power 
is required to reach the increasing threshold. 
Conversely, $\Iloss$ increases since, increasing $\rho_{\rm thr}$,
we are considering lower values of $f_{\rm thr}$ and the support $[f_{\rm thr},f_{\rm max}]$ increases.
From the arguments above
it follows that if $f_{\rm thr} \ll f_{5\%}^{\tLam}$, then $\Iloss \approx 1$ 
and the measurement of the tidal parameter
will be strongly affected by noise fluctuation
and by sensitivity limits,
with the possibility of an uninformative inference.

In App.~\ref{app:tidalinfo}, we apply the method discussed above
to the injections studied in Sec.~\ref{sec:analysis}, 
in order to prove that the injection studies are performed in an informative 
framework for the tidal parameter $\tLam$.

Using a GW170817-like template, i.e a waveform whose intrinsic and extrinsic parameters
are fixed to the maximum posterior probability of GW170817,
and setting $\rho_{\rm thr}=1$, 
we find $f_{\rm thr} \approx 800$~Hz for LIGO design sensitivity
and $f_{\rm thr}\approx 600$~Hz for Virgo design sensitivity,
while for a network of three detectors $f_{\rm thr} \approx 1$~kHz
and $\Iloss \approx 20\%$
\footnote{
	However, note that currently known events do not contain as much high-frequency information 
	as the signals displayed here. We will further discuss real GW events in Sec.~\ref{sec:gw170817}.
}.

\subsection{Systematic Errors}
\label{sec:syserr}

The probability
distribution of the data $d=h+n$ containing a signal $h$ and noise
is $\propto e^{-1/2(d-h|d-h)}$. Thus, the knowledge of $h$
at 1-$\sigma$ level is limited to a unit ball in Wiener space,
\be\label{eq:Wienerball}
(d-h|d-h)<1\,.
\ee
Systematic effects due to waveform modeling have been studied in
connection to Eq.~\eqref{eq:Wienerball}, e.g. \cite{Lindblom:2008cm,Damour:2010zb}.
Given a waveform model $h$ to approximate the {\it true} signal $s$ recorded in the data $d=s+n$
(where $n$ denotes the noise contribution),
and using the inequality $|h-d|\leq|h-s|+|s-d|=|\delta h|+|n|$ with
$\delta h = s-h$, Eq.~\eqref{eq:Wienerball} (or the condition $\sigma^2>1$) translates
into the criterion
\be
\label{eq:acc_crit}
(\delta h|\delta h) <\epsilon^2 \,,
\ee
with $\epsilon^2=1$ (or smaller, for a more strict requirement).
This equation corresponds to demanding that the systematics biases
become of the same order as the statistical ones when the noise level
is doubled~\cite{Damour:2010zb}.
It can be written in terms of the
faithfulness as [e.g. Eq.~(31) of~\cite{Damour:2010zb}]
\be 
\label{eq:LB}
\mathcal{F} > 1 - \frac{\epsilon^2}{2 \,\rho^2}\,,
\ee 
with $\epsilon^2\leq1$. Note that sometimes it is suggested to relax
this criterion by taking $\epsilon^2=N$, the number of intrinsic parameters of the
system \cite{Chatziioannou:2017tdw}.
The above criteria are necessary conditions that have to be satisfied
by faithful waveform models. Their violation does not guarantee
the presence of biases. Indeed in Sec.~\ref{sec:PEbiases} we show that all
of our simulated signals lie well below the faithfulness tresholds identified
by Eq.~\eqref{eq:LB}, though not all of them present obvious biases on $\tLam$.
Conversely, if a bias is present, they do not quantify how
large the uncertainty on the parameters is. 

The biases $\Delta \params = \bar{\params} - \params_{\rm true}$ between the
maximum likelihood (best fit) and the true parameters due to use of a waveform model $h$
instead of the exact waveform can be
estimated following~\cite{Cutler:2007mi,Favata:2013rwa}.
The best fit (possibly biased) values $\bar{\params}$ minimize the function $g(\params) = (d - h(\params)| d -h(\params))$. 
Therefore, they have to be
critical points of $g$, thus leading to the condition 
\be 
\label{eq:critical}
(\partial_{j} h(\bar{\params}) | d - h(\bar{\params})) = 0\,.
\ee
Linearly expanding $h(\bar{\params}) \approx h(\params_{\rm true}) + \Delta \params^j \partial_j h(\params_{\rm true})$
and inserting it in \eqref{eq:critical} one finds that 
\be 
\label{eq:bias}
\Delta \params^i = (F^{-1}(\params_{\rm true}))^{ij}(\partial_j h(\params_{\rm true}) | d - h(\params_{\rm true}))
\ee
This equation can be reconducted to the accuracy criterion of Eq.~\eqref{eq:acc_crit}. Indeed, recalling that
$\sigma^2_{ij} = (F^{-1})_{ij}$, we can write
\be 
(\sigma^{2}_{ij}(\params_{\rm true}))^{-1}\Delta \params^i = (\partial_j h(\params_{\rm true}) | d - h(\params_{\rm true}))
\ee 
multiplying both sides by $\Delta \params^j$, recalling that $h(\bar{\params}) \approx h(\params_{\rm true}) + \Delta\params^j \partial_j h(\params_{\rm true})$ and approximating $d \approx h(\bar{\params})$ immediately gives
\be  
\label{eq:bias_crit}
(\sigma^{2}_{ij} (\params_{\rm true}))^{-1}\Delta \params^i \Delta
\params^j \approx ( d - h(\params_{\rm true}) | d - h(\params_{\rm true}))
\ee 
Comparing Eq.~\eqref{eq:acc_crit} to Eq.~\eqref{eq:bias_crit} 
we note that indeed the validity of the former implies that the systematic biases $\Delta \params^i$ are smaller
than uncertainties due to statistical fluctuations, as expected.

Estimates of the parameters bias using Eq.~\eqref{eq:bias}
require knowledge of the derivatives of the waveform model with respect to the parameters.
These quantities, however, are  nontrivial to evaluate for more sophisticated semi-analytical approximants.
One might then try to directly minimize the function $g(\params)$. This in turn
requires the minimization of an integral in the multi-dimensional space
of the binary parameters, which can be computationally very expensive.
However, we are interested in the bias in the reduced tidal
parameter and thus assume that (i) the correlation with the other
parameters can be neglected; (ii) the largest biased parameter is $\tLam$.
The former assumption roughly holds if the SNR is sufficiently high
(see above); the latter if the point-mass waveforms are
sufficiently accurate at low frequencies.
In these conditions, minimizing the likelihood over the whole parameter space simply reduces to
computing
\be 
\label{eq:bias_min}
\min_{\tilde\Lambda}(d -h({\tilde\Lambda}) | d -h(\tLam))
\ee 
over a one-dimensional interval of $\tLam$ values, assuming that
all other intrinsic parameters are correclty estimated. 
While such a minimization has little practical use for GW PE, as the
true parameters $\params_{\rm true}$ are unknown, it can nonetheless be used
to estimate - known the parameters associated to one particular model -  the resulting value of $\tLam$ that one would get
by repeating PE with a different waveform model. Note the new 
model can disagree with the previous one also in the point-mass
and spin sector as assumption (ii) only requires the two models to agree in the low-frequency limit. 
In Sec.~\ref{sec:PEbiases} we apply this estimate to injection experiments. 
We find that it is able to correctly capture the behavior of the different
approximants studied, and that the estimated values of $\tLam$ 
(henceforth denoted as $\tLamB$) always fall within the 
$90\%$ credible intervals of the recovered posterior distributions, with the exception 
of few borderline cases where $\tLamB$ 
is nonetheless extremely close to the upper $95^{\rm th}$ percentile.
In Sec.~\ref{sec:3G}, instead, we apply Eq.~\eqref{eq:bias_crit} to two 
state-of-the-art approximants to estimate the importance of waveform systematics 
on PE with third generation detectors.

Note that the arguments presented in this section do not address the impact
of prior assumptions in GW PE, but rather focus on the maximum-likelihood
estimates, which exactly coincide with the maximum (posterior) probability values
only when considering uniform prior distributions.
As a general rule of thumb, as long as prior assumptions are more constraining on the source
parameters than the actual observational information carried by the waveform, one should expect
a-priori hypotheses to play an important role in PE~\cite{Favata:2013rwa}.
Extreme care is then required when dealing with lower SNR signals. For
example, as discussed in
\cite{Kastaun:2019bxo}, when sampling directly in the component tidal parameters
$\Lambda_A, \Lambda_B$ 
the prior on $\tLam$ is not independent 
of the mass ratio of the binary.
This, in turn, impacts the computation of credible bounds -- and especially of lower bounds, 
which are used to claim the measurement of tides. 
In the limit of high SNR, instead, the mean of the posterior distribution can be 
shown to coincide with the maximum likelihood estimators \cite{Vallisneri:2007ev}. Therefore,
it is in this regime that the discussion presented above has to be interpreted.

\section{Waveform models}
\label{sec:wf}

\begin{figure*}[t]
  \centering 
    \includegraphics[width=\textwidth]{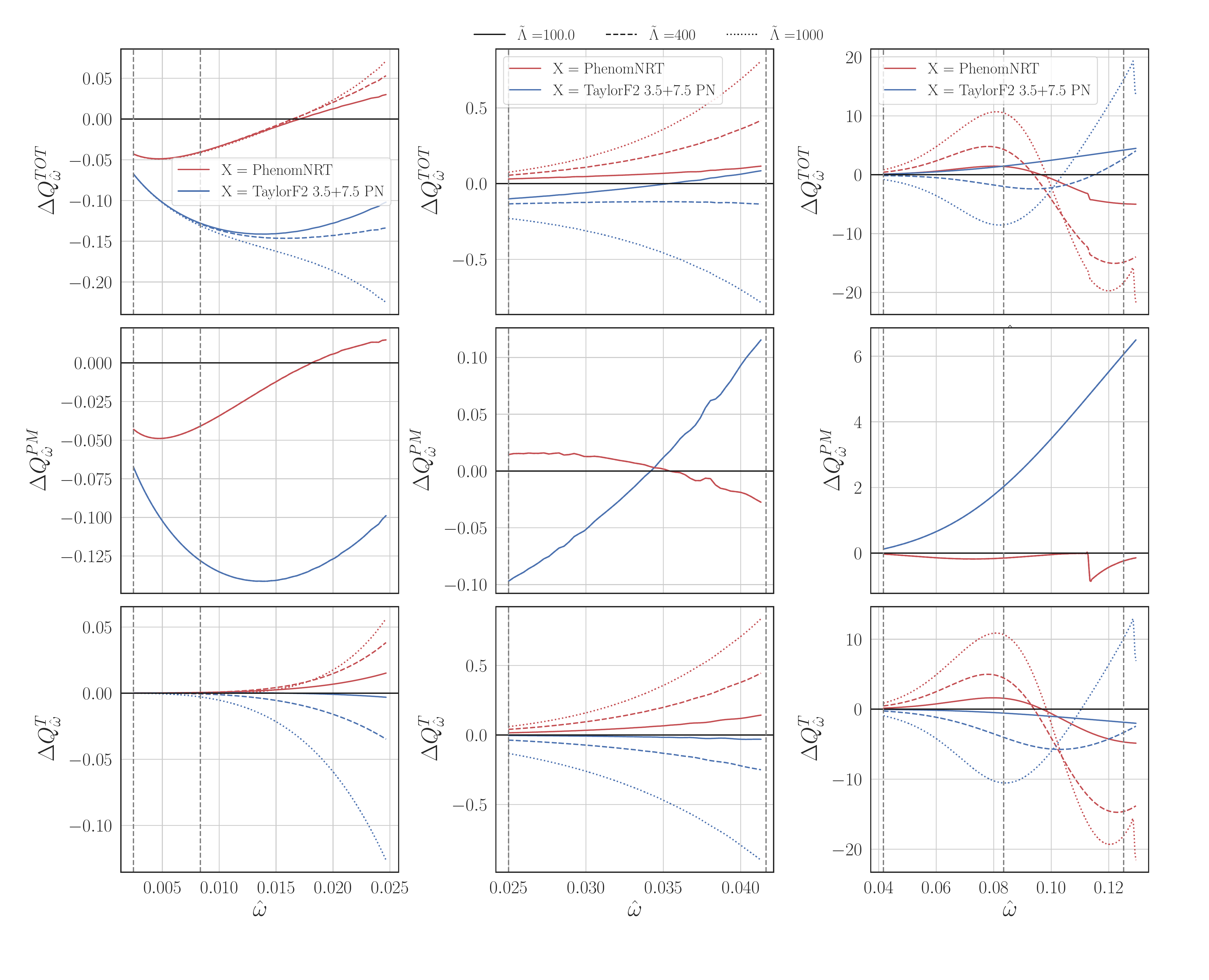}
    \caption{The $\Delta Q_{\omega} = Q_{\omega}^{TEOB} - Q_{\omega}^{X}$ function computed for three 
    waveforms with fixed spins $\chi_1 =\chi_2 = 0$ and varying  $\tLam = \{100, 400, 1000\}$,  represented by continuous,
    dashed and dotted lines, respectively. $\Delta Q_{\hat\omega}$ is then further decomposed into its point mass $\Delta Q_{\hat\omega}^{\rm PM}$ 
    (second row) and tidal $\Delta Q_{\hat\omega}^{\rm T}$ (third row) contributions, so that 
    $\Delta Q_{\hat\omega}^{\rm TOT} = \Delta Q_{\hat\omega}^{\rm PM} + \Delta Q_{\hat\omega}^{\rm T}$, 
    and is displayed over three different frequency ranges, roughly corresponding to the regimes in which point mass effects 
    are dominant, comparable or negligible with respect to tidal effects. We observe that $\Delta Q_{\hat\omega}^T$ for $\TF2$ 
    and $\NRT1$ have opposite behaviours, with $\TF2$ being more repulsive and $\NRT1$ more attractive than $\TEOB$.}
\label{fig:Qomg}
\end{figure*}

\begin{figure*}[t]
  \centering 
    \includegraphics[width=\textwidth]{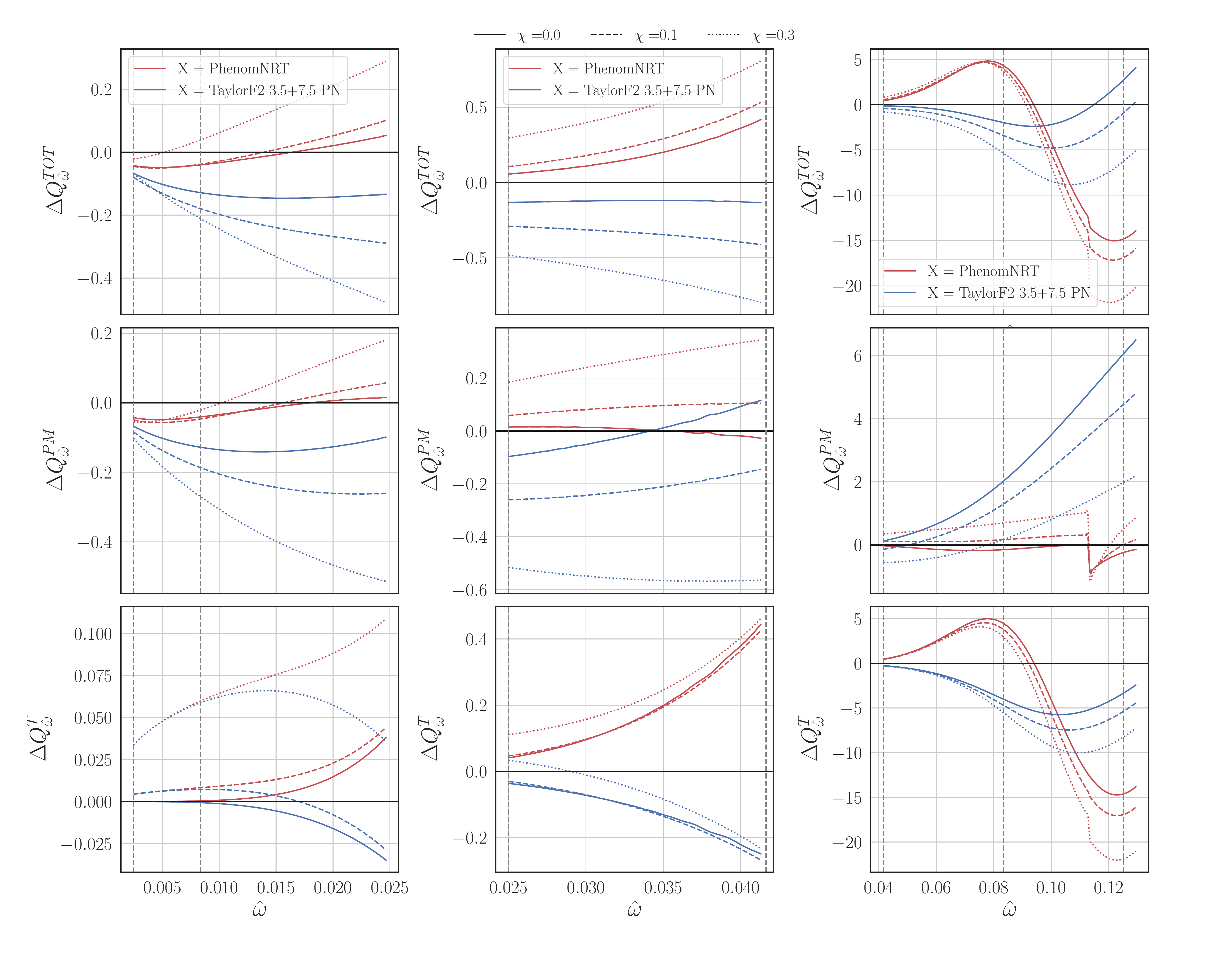}
    \caption{The $\Delta Q_{\hat\omega}^{\rm TOT} = Q_{\hat\omega}^{TEOB} - Q_{\hat\omega}^{X}$ function computed for three 
    waveforms with fixed $\tLam = 400$ and varying spins $\chi_1 =\chi_2 = \{0, 0.1, 0.3\}$, represented by continuous,
    dashed and dotted lines, respectively. $\Delta Q_{\hat\omega}$ is then further decomposed into its point mass $\Delta Q_{\hat\omega}^{\rm PM}$
    (second row) and tidal $\Delta Q_{\hat\omega}^{\rm T}$ (third row) contributions, 
    so that $\Delta Q_{\hat\omega}^{\rm TOT} = \Delta Q_{\hat\omega}^{\rm PM} + \Delta Q_{\hat\omega}^{\rm T}$. Note that $\Delta Q_{\hat\omega}^T$ is
    comparable to $\Delta Q_{\hat\Omega}^{\rm PM}$ at $\hat\omega < 0.02$. This effect can be attributed to the spin-spin interactions.}
    \label{fig:Qomg_spin}
\end{figure*}

Gravitational waveform models for coalescing compact binaries aim at providing approximate solutions 
to the GR two-body problem. They map a set of intrinsic parameters $\params$, for example the mass 
ratio $q$, the chirp mass $\mathcal{M}_c$, the component dimensionless spins $(\chi_A, \chi_B)$ 
and the dimensionless tidal deformabilities $(\Lambda_A, \Lambda_B)$, into
a time or frequency series $h(t; \params)$ or $\tilde{h}(f; \params)$.
Post newtonian (PN) approximants~\cite{Blanchet:2013haa,Buonanno:2009zt} 
construct this mapping by analytically computing the evolution of the orbital 
phase $\phi(t)$ of a binary system as a perturbative expansion in a 
small parameter $v/c$ or $x=(v/c)^2$, in which $v$ is the characteristic velocity 
of the binary. Such models, while cheap from a computational standpoint, are typically 
unable to reliably describe the waveform at high frequencies~\cite{Damour:1997ub}, 
i.e during the later phases of the evolution of the binary when $v$ becomes a comparable
fraction of $c$.
The effective-one-body (EOB) approach~\cite{Buonanno:1998gg, Buonanno:2000ef,Damour:2000we,Damour:2001tu,Damour:2008qf,Damour:2015isa,Bini:2019nra,Bini:2020wpo,Bini:2020nsb}
resums the PN information (both in the conservative and
nonconservative part of the dynamics) so to make it reliable and predictive also in
the strong-field, fast velocity regime. Once improved by NR data, this method allows 
one to compute the complete waveform from the early, quasi-adiabatic, inspiral up
to merger and -- when dealing with binary balck holes -- ringdown. 
Finally, phenomenological models~\cite{Santamaria:2010yb, Hannam:2013oca,Khan:2015jqa,Husa:2015iqa,London:2017bcn,Garcia-Quiros:2020qpx,Pratten:2020fqn,Pratten:2020ceb} are constructed by first stitching together EOB-based 
inspirals with numerical relativity simulations, when available, and then devising 
an accurate, effective, interpolating representation all over the parameter space 
devised to be computationally efficient.

For our purposes, we choose one representative approximant from each of the three families above. In particular, 
our analysis will employ the PN {\tt TaylorF2} model, the EOB {\tt
  TEOBResumS} model, and the Phenomenological {\tt
  IMRPhenomPv2NRTidal} model. In sections \ref{sec:3G} and Appendix~\ref{app:SelfSpin} we will 
then consider two further approximants:
{\tt IMRPhenomPv2NRTidalv2}
and {\tt SEOBNRv4Tsurrogate}.

{\tt TaylorF2} is a frequency domain PN waveform model. The phase of the GW, obtained through 
a stationary phase approximation, contains point-mass effects which are fully known up to
relative 3.5 PN order \cite{Buonanno:2009zt}, and include spin-spin and spin-orbit interactions~\cite{Arun:2008kb, Mikoczi:2005dn}.
A higher order, parameterized, quasi--5.5PN description of nonspinning point mass effects has also been derived in~\cite{Messina:2019uby}.
Tidal effects can be included up to relative 7.5PN order~\cite{Vines:2011ud, Damour:2012yf, Henry:2020ski},
while quadratic-in-spin effects were included up to 3.5PN~\cite{Nagar:2018zoe}.
Throughout the main body of this work we will employ a 3.5PN-accurate point mass baseline, 
a 7.5PN description of the tidal phasing, and a 3PN description of spin-square effects.

{\tt TEOBResumS} is a state-of-the-art EOB waveform model for spin-aligned 
coalescing compact binaries (either neutron stars or 
black holes)~\cite{Damour:2014sva,Nagar:2017jdw,Nagar:2018zoe,Nagar:2018plt,Nagar:2019wds,Nagar:2020pcj}.
In this paper, we focus on the tidal sector of {\tt TEOBResumS}, in the form
described in~\cite{Nagar:2018zoe,Akcay:2018yyh,Nagar:2018plt}.
In particular, this configuration coincides with the one implemented within 
{\tt LALInference}. The tidal sector of {\tt TEOBResumS} contains contributions from the
multipolar $\ell=2,3,4$ gravitoelectric and $\ell=2$ gravitomagnetic 
interations; the former are included in resummed form stemming from PN 
and gravitational-self force results~\cite{Bernuzzi:2015rla,Akcay:2018yyh} 
(see also Refs.~\cite{Bini:2012gu,Bini:2014zxa}). Equation of state-dependent
self-spin effects (also known as quadrupole-monopole terms) are included
at next-to-next-to-leading-order~\cite{Nagar:2018plt} thanks to a suitable
modification of the {\it centrifugal radius} introduced in Ref.~\cite{Damour:2014sva},
so to incorporate even-in-spin effects in a way that closely mimics the structure
of the Hamiltonian of a point-particle on a Kerr black hole.
In addition, the models relies on the (iterated) post-adiabatic 
approximation~\cite{Damour:2012ky,Nagar:2018gnk,Akcay:2018yyh} 
to  compute the full inspiral waveform until about 10 orbits before merger, 
so to greatly reduce the computational burden of the waveform generator 
with negligible losses of accuracy.

These choices, together with rather different treatment of the spin sector and
of resummation choices  distinguish {\tt TEOBResumS} from the other state of the 
art EOB approximant, {\tt SEOBNR} \cite{Pan:2013tva, Babak:2016tgq}. We address
the reader to Ref.~\cite{Rettegno:2019tzh} for a detailed investigation of the differences 
between the conservative point-mass dynamics of the models. To improve the computational efficiency of the waveform generation,
when considering BNS systems, the {\tt SEOBNR} family applies gaussian process regression
to the baseline model {\tt SEOBNRv4T}~\cite{Hinderer:2016eia, Steinhoff:2016rfi, Bohe:2016gbl} -- which includes a description of dynamical 
tides, but no self-force information -- so to obtain {\tt SEOBNRv4Tsurrogate}~\cite{Lackey:2018zvw}.
Note that EOB models are the most analytically complete to date, and contain 
higher order PN information than that contained in Taylor expanded PN 
approximants (e.g. many more test-particle terms at higher PN order as 
well as resummed tail factor). For this reason, the EOB framework can be 
Taylor-expanded so to obtain waveform approximants at (partial) higher PN order than
the currently, fully known, 3.5PN one~\cite{Damour:2012yf, Messina:2017yjg, Messina:2019uby}.

{\tt IMRPhenomPv2NRTidal} is a phenomenological spin precessing model
for BNS systems based on the {\tt IMRPhenomPv2} model. In the latter, 
an effective description of the point-mass waveform is obtained by fitting {\tt SEOBNR}-NR 
hybrid waveforms\footnote{These waveforms are obtained by stitching together inspiral
waveforms for the long inspiral to NR simulations that go through merger and ringdown.} 
to an analytical representation of the amplitude and phase of the frequency 
domain 22 mode $h_{22}$ \cite{Husa:2015iqa, Khan:2015jqa}. This representation 
is further augmented by the {\tt NRTidal} model \cite{Dietrich:2017aum}, 
which provides a description of tidal effects  based on a fit of hybrid 
waveforms composed of PN, {\tt TEOBResumS}
and nonspinning, $q \approx 1$ NR simulations. 

Recently, Ref.~\cite{Dietrich:2019kaq} improved this model to
{\tt IMRPhenomPv2NRTidalv2} by incorporating a 7.5PN-accurate 
low frequency limit for the tidal sector of the phasing and 
PN-expanded spin-quadrupole interactions up to 3.5PN 
in the waveform phase together with new fits for 
the amplitude tidal corrections. In this work, we will use
both {\tt IMRPhenomPv2NRTidal} and {\tt IMRPhenomPv2NRTidalv2}
imposing that the individual spins are aligned to the orbital angular 
momentum.

%===================================
\subsection{Comparing waveform approximants}
\label{sec:comparing_waveforms}
%===================================
Let us now turn to discussing in some detail how the
differences in the approximants reflect on the GW phase.
This is the very first step to take towards the 
understanding of waveform systematics. Given the plus 
and cross polarizations $h_{+}(t, \params), h_{\times}(t, \params)$ associated to a specific approximant,
we define the frequency domain waveform
$\tilde{h}(f, \params) = \tilde{A}(f, \params) e^{-i \Psi(f,  \params)}$, 
where $\tilde{A} = |\tilde{h}_{+}(f) + i \tilde{h}_{\times}(f)|$, $\Psi(f) = -{\rm arg}(\tilde{h}_{+}(f) + i \tilde{h}_{\times}(f))$ and $\tilde{h}_{+, \times}$ 
are the Fourier transforms of the time domain polarizations. 
Extracting information directly from the waveform phasing $\Psi(f)$ is complicated
by the presence of an affine linear term $\phi_c + 2\pi t_c f$ which
can be fixed arbitrarily.
A a better quantity to discuss waveform phasing is 
\be
\label{eq:qomg}
Q_{\hat{\omega}} = \frac{\hat{\omega}^2}{\dot{\hat{\omega}}} =
\frac{d \phi(t)}{d\ln\hat\omega}
\ee
where $\hat{\omega} = 2 \pi f M$ is the dimensionless GW
frequency. The time-domain GW phase accumulated between two
frequencies is given by 
\be
\phi_{(\omega_1,\omega_2)} = \int_{\hat\omega_1}^{\hat\omega_2} Q_{\hat\omega} d \ln\hat\omega \ .
\ee
Physically, $Q_{\hat{\omega}}$ is related to the phase acceleration,
and the GW phase in the Stationary Phase Approximation (SPA) is given
by $\Psi''(w)=Q_{\hat{\omega}}(w)/w^2$. 
The inverse of $Q_{\hat\omega}$ is thus the adiabatic parameter whose
magnitude controls the validity of the SPA \cite{Baiotti:2011am,Nagar:2018zoe,Messina:2019uby}. 
Since there is no time/phase shift ambiguity and no necessity of alignment in 
phase plots with the $Q_{\hat{\omega}}$, the latter quantity is
preferable with respect to the phase because information can be lost
in the alignement~\cite{Baiotti:2010xh,Baiotti:2011am,Damour:2012ky}.
Thus rather, than comparing phase differences, we compute $Q_{\hat\omega}$ for the
waveform approximants discussed above, and extract information from
$\Delta Q_{\hat\omega}= Q_{\hat\omega}^{\TEOB} - Q_{\hat\omega}^{X}$, where $X$ is
any other approximant.

Figure~\ref{fig:Qomg} shows the quantity $\Delta Q_{\hat\omega}$, 
computed for three reference waveforms with varying $\tLam$ and zero spins and decomposed
into its point-mass $\Delta Q_{\hat{\omega}}^{PM}$
and tidal $\Delta Q_{\hat{\omega}}^{T}$ contributions. 
The frequency range is roughly divided at the ``cutoff'' thresholds
of the regimes at which point mass ($\hat{\omega}<0.02$) and tidal
($\hat{\omega}>0.05$) effects are measured according to the
  Fisher matrix information formalism.
During the early inspiral (first column), point mass contributions dominate over tidal effects, 
and as expected the phenomenological description of the inspiral is closer to {\tt TEOBResumS} than the one offered by {\tt TaylorF2}.
When $0.02\lesssim \hat{\omega}\lesssim 0.05$ (second column) the importance of tidal 
effects gradually increases, and the behavior of the two 
approximants starts differing significantly.
Focusing on {\tt IMRPhenomPv2NRTidal}, we observe that the largest contribution to 
$\Delta Q_{\hat\omega}$  comes from the tidal sector. As $\tLam$
grows, both $\Delta Q_{\hat\omega}^T$ and $\Delta Q_{\hat\omega}$ become 
increasingly more positive. Therefore, matter effects in {\tt IMRPhenomPv2NRTidal} 
are stronger than in {\tt TEOBResumS}.
Over the same range ($ 0.02 \lesssim \hat{\omega} \lesssim 0.05$), tidal terms of {\tt TaylorF2} 
behave in the exact opposite way. Increasing the value of $\tLam$ leads to more negative 
$\Delta Q_{\hat\omega}^T$. 
For this approximant, then, matter effects are weaker than {\tt TEOBResumS}.
The trends shown in the intermediate range are maintained by both approximants also for $\hat{\omega} > 0.05$ 
and up to $\hat{\omega} \approx 0.10$, close to merger frequency (third column). We highlight that the point mass
terms of {\tt TaylorF2} grow monotonically, reflecting how the PN approximation breaks down at high frequencies. 
However, notably, the point mass contribution is positive -- i.e more attractive than 
{\tt TEOBResumS}' -- and larger than or 
comparable to tidal corrections for moderate values of
$\tLam$. In GW parameter estimation, $\Delta Q_{\hat\omega}^{PM}$ 
then can partially compensate the negative $\Delta Q_{\hat\omega}^T$.
Globally, 
{\tt IMRPhenomPv2NRTidal} is more attractive than {\tt TEOBResumS}, which implies that when 
recovering simulated {\tt TEOBResumS} waveforms with {\tt IMRPhenomPv2NRTidal}, one may expect 
to find lower values of $\tLam$ than the ones injected.
Instead, when recovering simulated {\tt TEOBResumS} waveforms with {\tt TaylorF2}, one may expect 
to find higher values of $\tLam$ than the ones injected.

Spin effects are studied with a similar approach in
Fig.~\ref{fig:Qomg_spin}, which shows $\Delta Q_{\hat\omega}$ computed for three waveforms with fixed 
$\tLam=400$ and varying magnitude of the dimensionless spins $(\chi_A,\chi_B)$. 
We consider configurations with spins aligned to the orbital 
angular momentum and such that $\chi_A=\chi_B = \chi$.
Focusing on the point mass contribution, we observe that increasing $\chi$ does not impact significantly 
the magnitude of $\Delta Q_{\hat\omega}^{PM}$ for {\tt IMRPhenomPv2NRTidal}. On the other hand, spin-induced 
effects are noticeably more repulsive in {\tt TaylorF2} than in {\tt TEOBResumS} over the whole frequency range considered.
Concerning $\Delta Q_{\hat\omega}^T$, we observe that the differences at $\hat{\omega} < 0.02$ are no 
longer negligible with respect to the point mass contributions, and in general are larger than those 
found for non-spinning binaries. These differences can be attributed to the modelization 
of the spin-quadrupole terms. We recall that a spinning NS acquires a quadrupole  moment due to
its own rotation, which in turn causes a distortion of the gravitational field
outside the body. The magnitude of such quadrupole moment is an equation of state-dependent quantity, which can be parameterized
through a coefficient $C_Q$ \cite{Poisson:1997ha,Nagar:2018plt}. The importance of this term in parameter 
estimation was pointed out in e.g \cite{Harry:2018hke}, which showed how neglecting it can lead to biases 
on the recovery of the mass ratio and the total mass. 
Both {\tt TaylorF2} and {\tt IMRPhenomPv2NRTidal} include these corrections only up to 3PN (NLO),
 whereas {\tt TEOBResumS} also incorporates tail-dependent corrections in resummed form,
 as well as NNLO effects.  The resummation weakens 
the effect of quadrupole-monopole terms above $\hat{\omega} \approx 0.06$~\cite{Nagar:2018zoe,Nagar:2018plt}, 
i.e. above the frequency at which the NSs enter in contact and
hydrodynamical effects become relevant~\cite{Bernuzzi:2012ci}.
Note the weaker effect of the (effective) EOS-dependent self-spin terms with respect to
the PN expressions at high frequencies is also suggested by NR
simulations \cite{Dietrich:2016lyp}, with the latter also suggesting stronger
(effective) spin-orbit effects then PN~\footnote{%
  But note that in hydrodynamical regime it is, strictly speaking, not possible to
  interpret these as spin-interactions and to compare to PN.}.

Overall, when considering  injections of {\tt TEOBResumS} highly spinning waveforms we expect {\tt IMRPhenomPv2NRTidal} to underestimate
tidal parameters, and {\tt TaylorF2} to overestimate them.

\section{Injection Study}
\label{sec:analysis}

\begin{table*}[t]
\caption{Comparison between the properties of the injected signals and 
the recovered marginalized one dimensional posteriors. For each simulation
we report medians and 90\% credible regions. For each approximant 
and frequency range we additionally display the values of $\tLamB$ obtained as 
described in Sec.~\ref{sec:PEbiases}.}
\label{tab:posteriors}
\resizebox{\textwidth}{!}{\begin{tabular}{cccc|cccc|cccc|cccc|cccc}
\hline  \hline
 & \multicolumn{3}{c|}{Injection}&\multicolumn{8}{c|}{{\tt
    IMRPhenomPv2\_NRT}} & \multicolumn{8}{c}{{\tt TaylorF2}
  (3.5PN+7.5PN tides)}\\
\hline \hline
$f_{\rm cut}$& \multicolumn{3}{c|}{}&\multicolumn{4}{c|}{1 kHz} & \multicolumn{4}{c|}{2 kHz} & \multicolumn{4}{c|}{1 kHz}& \multicolumn{4}{c}{2 kHz}\\
\hline\hline
EOS & $M_{\rm inj}$ & $q_{\rm inj}$ & $\tLam_{\rm inj}$ & $M$ & $q$ & $\tLam$ & $\tLamB$& $M$ & $q$ & $\tLam$ & $\tLamB$ & $M$ & $q$ & $\tLam$ &  $\tLamB$ & $M$ & $q$ & $\tLam$ &  $\tLamB$\\
\hline  
DD2 & $2.71$ & $1.00$ & $840$ & $2.70^{+0.06}_{-0.02}$ & $0.83^{+0.15}_{-0.18}$ & $628^{+130}_{-144}$ & 758 & $2.71^{+0.06}_{-0.03}$ & $0.78^{+0.19}_{-0.15}$ & $622^{+93}_{-112}$ & 711 & $2.70^{+0.06}_{-0.02}$ & $0.83^{+0.15}_{-0.18}$ & $916^{+188}_{-192}$ & 918 & $2.70^{+0.06}_{-0.02}$ & $0.84^{+0.15}_{-0.18}$ & $1011^{+144}_{-152}$ & 920\\
LS220 & $2.68$ & $1.00$ & $715$ & $2.68^{+0.06}_{-0.02}$ & $0.83^{+0.15}_{-0.19}$ & $525^{+135}_{-140}$ & 660 & $2.68^{+0.06}_{-0.02}$ & $0.79^{+0.18}_{-0.16}$ & $528^{+87}_{-104}$ & 613 & $2.68^{+0.06}_{-0.02}$ & $0.82^{+0.16}_{-0.18}$ & $758^{+188}_{-195}$ & 754 & $2.67^{+0.06}_{-0.02}$ & $0.84^{+0.15}_{-0.18}$ & $844^{+145}_{-144}$ & 758\\
LS220 & $2.69$ & $0.86$ & $714$ & $2.68^{+0.07}_{-0.02}$ & $0.81^{+0.16}_{-0.18}$ & $532^{+137}_{-147}$ & 633 & $2.69^{+0.07}_{-0.03}$ & $0.78^{+0.19}_{-0.16}$ & $532^{+99}_{-108}$ & 589 & $2.68^{+0.07}_{-0.02}$ & $0.81^{+0.17}_{-0.19}$ & $764^{+191}_{-195}$ & 756 & $2.68^{+0.06}_{-0.02}$ & $0.82^{+0.16}_{-0.18}$ & $856^{+149}_{-158}$ & 756\\
SFHo & $2.71$ & $1.00$ & $413$ & $2.70^{+0.06}_{-0.01}$ & $0.84^{+0.15}_{-0.18}$ & $293^{+118}_{-117}$ & 388 & $2.70^{+0.06}_{-0.02}$ & $0.83^{+0.15}_{-0.18}$ & $306^{+75}_{-84}$ & 352 & $2.71^{+0.07}_{-0.02}$ & $0.82^{+0.16}_{-0.18}$ & $409^{+174}_{-169}$ & 453 & $2.71^{+0.07}_{-0.02}$ & $0.82^{+0.16}_{-0.18}$ & $443^{+124}_{-125}$ & 449\\
SFHo & $2.72$ & $0.88$ & $412$ & $2.70^{+0.06}_{-0.02}$ & $0.83^{+0.15}_{-0.19}$ & $299^{+119}_{-119}$ & 341 & $2.70^{+0.06}_{-0.02}$ & $0.82^{+0.16}_{-0.18}$ & $304^{+76}_{-82}$ & 350 & $2.71^{+0.07}_{-0.02}$ & $0.82^{+0.17}_{-0.19}$ & $416^{+172}_{-166}$ & 440 & $2.71^{+0.07}_{-0.02}$ & $0.81^{+0.17}_{-0.18}$ & $439^{+132}_{-131}$ &  435\\
SLy & $2.68$ & $1.00$ & $401$ & $2.67^{+0.06}_{-0.01}$ & $0.84^{+0.14}_{-0.18}$ & $286^{+120}_{-120}$ & 349 & $2.67^{+0.06}_{-0.01}$ & $0.84^{+0.14}_{-0.18}$ & $295^{+72}_{-80}$ & 350 & $2.68^{+0.07}_{-0.02}$ & $0.83^{+0.15}_{-0.19}$ & $397^{+173}_{-169}$ & 434 & $2.68^{+0.07}_{-0.02}$ & $0.82^{+0.16}_{-0.19}$ & $423^{+122}_{-124}$ & 430\\
SLy & $2.69$ & $0.88$ & $401$ & $2.68^{+0.06}_{-0.02}$ & $0.83^{+0.15}_{-0.18}$ & $290^{+123}_{-124}$ & 358 & $2.68^{+0.06}_{-0.02}$ & $0.83^{+0.16}_{-0.18}$ & $294^{+77}_{-85}$ & 313 & $2.68^{+0.07}_{-0.02}$ & $0.81^{+0.17}_{-0.18}$ & $404^{+176}_{-169}$ & 557 & $2.68^{+0.07}_{-0.02}$ & $0.81^{+0.17}_{-0.18}$ & $426^{+127}_{-125}$ & 386\\
DD2 & $2.48$ & $1.00$ & $1366$ & $2.47^{+0.05}_{-0.02}$ & $0.83^{+0.15}_{-0.16}$ & $1104^{+186}_{-189}$ & 1269 & $2.48^{+0.04}_{-0.02}$ & $0.78^{+0.19}_{-0.12}$ & $1057^{+141}_{-138}$ & 1170 & $2.47^{+0.04}_{-0.01}$ & $0.84^{+0.14}_{-0.17}$ & $1542^{+246}_{-245}$ & 1477 & $2.47^{+0.04}_{-0.01}$ & $0.85^{+0.13}_{-0.17}$ & $1668^{+200}_{-207}$ & 1484\\
DD2 & $3.18$ & $1.00$ & $332$ & $3.17^{+0.07}_{-0.02}$ & $0.84^{+0.14}_{-0.18}$ & $218^{+73}_{-76}$ & 257 & $3.18^{+0.07}_{-0.02}$ & $0.81^{+0.17}_{-0.16}$ & $248^{+55}_{-61}$ & 297 & $3.17^{+0.08}_{-0.02}$ & $0.82^{+0.16}_{-0.18}$ & $311^{+118}_{-115}$ & 345 & $3.18^{+0.08}_{-0.03}$ & $0.81^{+0.17}_{-0.18}$ & $335^{+96}_{-99}$ & 343\\
2B & $2.70$ & $1.00$ & $127$ & $2.69^{+0.05}_{-0.01}$ & $0.86^{+0.12}_{-0.18}$ & $115^{+105}_{-76}$ & 116 & $2.69^{+0.05}_{-0.01}$ & $0.86^{+0.13}_{-0.18}$ & $98^{+52}_{-52}$ & 117 & $2.69^{+0.06}_{-0.01}$ & $0.84^{+0.15}_{-0.18}$ & $150^{+142}_{-106}$ & 136 & $2.69^{+0.06}_{-0.02}$ & $0.83^{+0.15}_{-0.18}$ & $106^{+92}_{-72}$ & 130\\
SLy & $3.00$ & $1.00$ & $191$ & $2.99^{+0.06}_{-0.01}$ & $0.85^{+0.13}_{-0.18}$ & $132^{+81}_{-74}$ & 168 & $2.99^{+0.06}_{-0.02}$ & $0.85^{+0.13}_{-0.18}$ & $143^{+49}_{-53}$ & 169 & $2.99^{+0.07}_{-0.02}$ & $0.83^{+0.15}_{-0.18}$ & $169^{+124}_{-105}$ & 307 & $3.00^{+0.07}_{-0.02}$ & $0.81^{+0.17}_{-0.18}$ & $158^{+87}_{-84}$ & 162\\
LS220 & $3.20$ & $1.00$ & $202$ & $3.19^{+0.06}_{-0.02}$ & $0.85^{+0.13}_{-0.18}$ & $132^{+66}_{-67}$ & 158 & $3.19^{+0.07}_{-0.02}$ & $0.85^{+0.14}_{-0.18}$ & $156^{+47}_{-54}$ & 158 & $3.19^{+0.08}_{-0.02}$ & $0.83^{+0.15}_{-0.18}$ & $168^{+108}_{-96}$ & 288 & $3.20^{+0.08}_{-0.03}$ & $0.81^{+0.17}_{-0.18}$ & $171^{+86}_{-83}$ & 274\\
SFHo & $2.92$ & $1.00$ & $252$ & $2.91^{+0.06}_{-0.02}$ & $0.85^{+0.14}_{-0.18}$ & $172^{+90}_{-84}$ &  222 & $2.91^{+0.06}_{-0.02}$ & $0.84^{+0.14}_{-0.18}$ & $185^{+56}_{-59}$ & 224 & $2.92^{+0.07}_{-0.02}$ & $0.83^{+0.15}_{-0.18}$ & $233^{+135}_{-127}$ & 257 & $2.91^{+0.07}_{-0.02}$ & $0.81^{+0.17}_{-0.18}$ & $233^{+96}_{-97}$ & 253\\
SFHo & $2.80$ & $1.00$ & $334$ & $2.79^{+0.06}_{-0.01}$ & $0.84^{+0.14}_{-0.18}$ & $228^{+106}_{-103}$ & 285 & $2.79^{+0.06}_{-0.01}$ & $0.84^{+0.14}_{-0.18}$ & $248^{+71}_{-75}$ & 286 & $2.79^{+0.07}_{-0.02}$ & $0.83^{+0.16}_{-0.19}$ & $319^{+156}_{-153}$ & 462 & $2.79^{+0.07}_{-0.02}$ & $0.81^{+0.17}_{-0.18}$ & $342^{+113}_{-118}$ & 444\\
ALF2 & $3.00$ & $1.00$ & $382$ & $2.99^{+0.07}_{-0.02}$ & $0.84^{+0.14}_{-0.18}$ & $258^{+88}_{-88}$ & 330 & $3.00^{+0.07}_{-0.02}$ & $0.81^{+0.16}_{-0.17}$ & $280^{+68}_{-72}$ & 330 & $2.99^{+0.07}_{-0.02}$ & $0.83^{+0.16}_{-0.19}$ & $370^{+138}_{-133}$ & 467 & $3.00^{+0.07}_{-0.03}$ & $0.81^{+0.17}_{-0.17}$ & $400^{+109}_{-110}$ & 457\\
\hline \hline
\end{tabular}}
\end{table*}

We present a full Bayesian PE study on 15 signals injected with
$\TEOB$ and recovered with $\TF2$ and $\NRT1$.
We interpret our results in light of the $Q_{\omega}$ analysis of Sec.~\ref{sec:wf},
and find that it correctly indicates the behaviors of the studied wavaform approximants. 

\subsection{Method}
In order to study waveform systematics in a controlled environment,
we generate artificial data strains ({\it injections}) using the {\tt TEOBResumS} model
(with all higher modes up to $\ell = 8$) for 15 different nonspinning binary configurations, 
described by the intrinsic parameters 
$(m_A, m_B, \Lambda_A,\Lambda_B)$ and reported in Table~\ref{tab:posteriors}
with the alternative representation $(M, q, \tilde\Lambda)$.
The waveform polarizations are then projected
on the three LIGO-Virgo detectors, locating the source at the sky position of GW170817~\footnote{We use the maximum posterior values for sky location and distance
	 from LVC analysis~\cite{TheLIGOScientific:2017qsa} combined with the information coming from \cite{GBM:2017lvd}.}.
The injections are 64~s long with a sampling rate of 4096~Hz
and they are performed with zero-noise configuration,
i.e. no additional noise is included in the analyzed strains,
in order to minimize the statistical fluctuations and to 
work in a framework as close as possible to the one described in Sec.~\ref{sec:ana}.
We use Advanced LIGO and Advanced 
Virgo design amplitude spectral densities (ASD)~\cite{Harry:2010zz,Aasi:2013wya,TheLIGOScientific:2014jea,TheVirgo:2014hva}.
The SNRs of the injected signals span a range
from 82 to 94 (depending on the specific combination of masses and tidal parameters), 
that result in louder signals than the current BNS observations~\cite{LIGOScientific:2018mvr, Abbott:2020uma}.
For the estimation of the posterior distributions, 
we adopt the Bayesian framework 
offered by the {\tt lalinference\_mcmc} sampler as implemented 
in the software LSC Algorithm Library Suite 
({\tt LALSuite})~\cite{lalsuite,Veitch:2009hd,Veitch:2014wba}.
The waveform models used in the matched filtering analysis are 
the already described $\TF2$ and $\NRT1$.

We perform two sets of injections, in part already discussed in~\cite{Agathos:2019sah}.
In the first set, matter effects are modeled using two
independent quadrupolar tidal parameters $\Lambda_A, \Lambda_B$.
In the second set, we use the spectral parametrization of the EOS~\cite{Lindblom:2010bb,Carney:2018sdv,Abbott:2018exr}. 
Within this framework,
the EOS of cold dense NS matter is represented as a smooth function,
parametrized in a 4-dimensional space by the coefficients 
$(\gamma_{0},\gamma_{1},\gamma_{2},\gamma_{3})$.
Each combination of these values specifies an 
adiabatic index $\Gamma(P)$
\begin{equation}
\label{eq:Gamma_spec}
\Gamma(P) =  \exp
\left[
\sum_{k=0}^{3} \gamma_{k} \log (P/P_{0})^{k}
\right] \, ,
\end{equation}
where $P_{0}$ is some reference pressure.
The adiabatic index is by definition related to 
the pressure-density function $P(\rho)$
through $\Gamma = \rho \frac{\dee\ln P}{\dee\rho}$.
The complete EOS is then built by fixing the low-density sector
$(P < P_0)$ to the SLy description, and integrating the differential equation
for $\rho(P)$ implied by the definition of $\Gamma$ in the core of the NS $(P > P_0)$.
Once the EOS is fixed, it is possible to calculate
the tidal polarizability parameters $\Lambda_{A,B}$,
which are then used to model the tidal effects in
the waveforms. These analyses give a posterior distribution for 
the coefficients $\gamma_i$, which can be mapped into EOSs and radii 
of the merging NSs. 
However, this method assumes implicitly that both NSs are 
described by the same EOS and that no strong first-order phase 
transitions happen in the core of the NS. 

For both the previous methods, the analyses are performed with two 
different maximal frequencies, $f_{\rm max}=1~{\rm kHz}$ and 
$f_{\rm max}=2~{\rm kHz}$, i.e. in frequency ranges $f\in[23~{\rm Hz}, 1~{\rm kHz}]$ and $f\in[23~{\rm Hz}, 2~{\rm kHz}]\,$, 
in order to verify if the extension to the higher frequency cutoff introduces additional biases.  
The priors distributions are flat in mass components,
in a range corresponding to 
$\mathcal{M}_c\in [1.0 , 2.2]~\Mo$ and $q\in [1,8]$.
We use aligned-spin configuration with 
isotropic priors on the spin components 
and $a_{i,z}\in [-0.05 , +0.05]$, $i=A,B$. 
Regarding the tidal parameters, 
the prior distributions are uniform in the free parameters involved in the analysis:
when we adopt the EOS-insensitive description,
$p(\Lambda_i)\propto 1 $ in the range $\Lambda_i\in[0 , 5000]$ for $i=A,B$; 
while for the spectral parametrization cases, 
the prior distribution is uniform  in the spectral parameters
in the ranges $\gamma_0 \in [0.2,2]$, 
$\gamma_1 \in [-1.6,1.7]$, 
$\gamma_2 \in [-0.6,0.6]$, 
$\gamma_3 \in [-0.02,0.02]$,
and additionally $\Gamma(P)$ is constrained to be in the range $[0.5,4.5]$.
This setup is identical to the one proposed in Ref.~\cite{Abbott:2018exr}.
In comparison to previous studies we employ a larger set of simulated signals, 
in order to better understand the behavior of the studied approximants when different sources are
considered~\cite{Dudi:2018jzn,Samajdar:2018dcx}. 

In the remainder of this section, we (i) examine
the measurement of mass and spin parameters, (ii) discuss the systematic
effects that different approximants induce in the recovered tidal  
parameters, NS radii and EOS reconstruction and (iii) apply the faithfulness
criteria previously described to our data.
%==========================

%==========================
\begin{figure}[t]
  \centering 
    \includegraphics[width=0.49\textwidth]{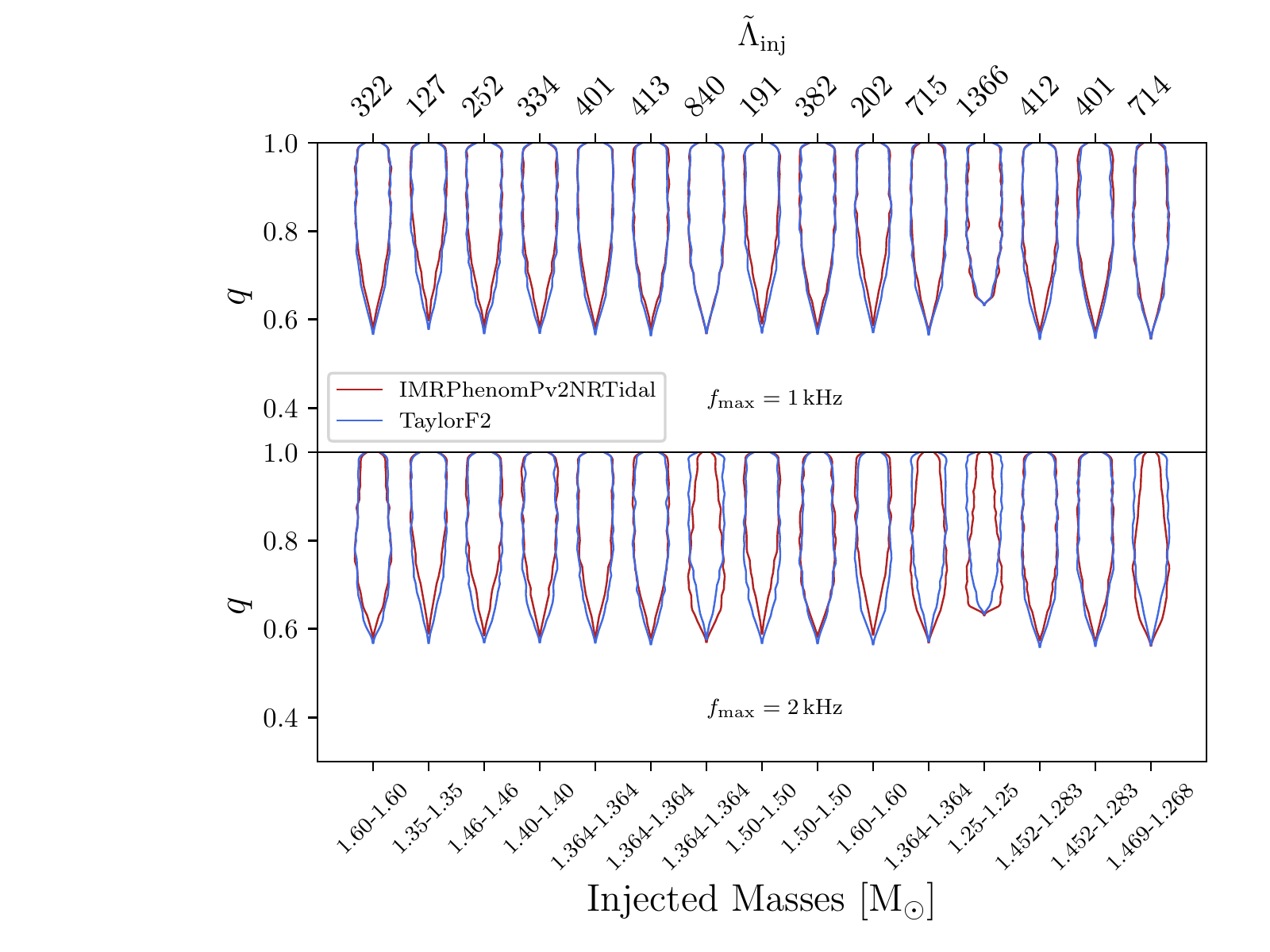}
    \includegraphics[width=0.49\textwidth]{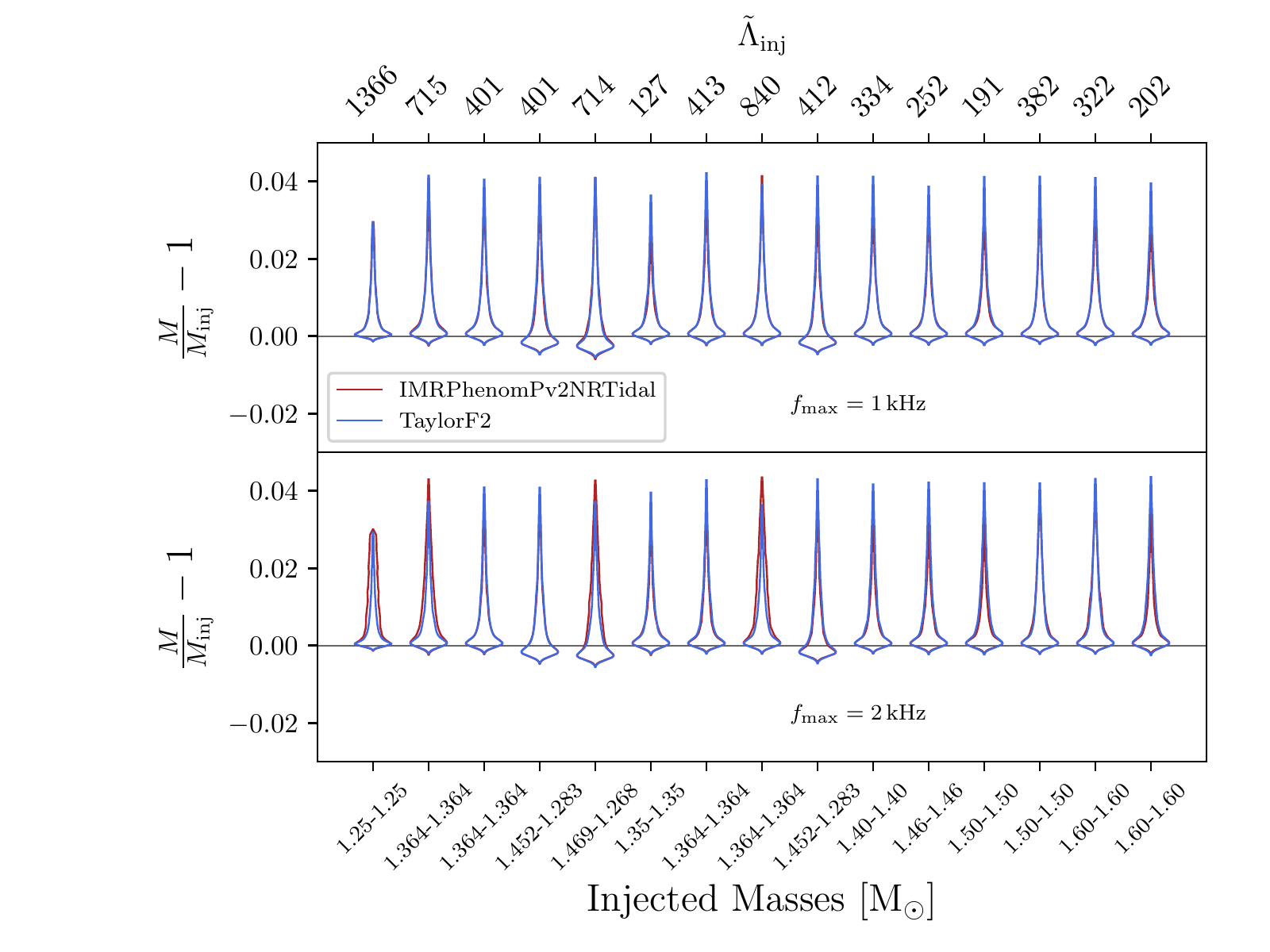}
    \caption{Distributions of the mass ratio $q$ (top) and the deviation from the injected total mass 
    $M_{\rm inj}$ (bottom), displayed for all the simulated signals of Table~\ref{tab:posteriors}.
    The distributions recovered are consistent between different approximants and frequency ranges. The 
    unequal mass signals cannot be distinguished from their equal mass counterpart.}
 \label{fig:mass_violin}
\end{figure}

\subsection{Masses, mass-ratio and spins}
\label{sec:nontidal}
We first discuss the determination of the nontidal parameters.
Figure~\ref{fig:mass_violin} shows the recovered posterior distributions of the  total mass $M$
and mass ratio $q$ parameters obtained with low spin priors. 
The estimates obtained are consistent between the different approximants, frequency cutoffs and with
the injected signals, with the real values always falling inside the $90\%$ credible intervals.
This indicates that the systematic differences in phasing observed at low frequencies (see the 
first column of Fig.~\ref{fig:Qomg}) are smaller than statistical uncertainties.  
We find that the injected unequal-mass signals (with $q = 0.86$ and $q = 0.88$) cannot be distinguished 
from the equal mass ones. 
This can partly be attributed to the known existing correlation between mass 
ratio and spin parameters~\cite{Cutler:1994ys}. In PN waveforms the leading order spin interactions 
are described by the parameter $\beta$, given by \cite{Cutler:1994ys, Arun:2008kb}
\be
\beta = \chi_{\rm eff} - \frac{38 \nu}{113}(\chi_1 + \chi_2), \label{eq:beta}
\ee
where
\be
\chi_{\rm eff} = \frac{m_1 \bm{\chi}_1 + m_2 \bm{\chi}_2}{M} \cdot \bm{\hat{L}}\label{eq:chi_eff} 
\ee
is the mass weighted sum of the component spin parameters, and is often times used during PE as a measure
of the collective spin of the binary, as it is a conserved quantity of the orbit-averaged precession
equations over precession timescales~\cite{Racine:2008qv}.
A Fisher Matrix analysis reveals that spin parameters, which at leading order have $p=-4$ in the notation
 of Sec.~\ref{sec:ana}, are measured over a very similar range of 
frequencies as the (symmetric) mass ratio $\nu$~\cite{Damour:2012yf, Harry:2018hke},
 to which they are therefore strongly correlated. 
In more detail, positive aligned spins have a repulsive effect on the binary dynamics. 
By contrast, decreasing the symmetric mass ratios (i.e, more unequal-mass systems) 
accelerates the coalescence. 
The two effects are thus in direct competition, and spin effects can be reproduced by
varying $\nu$~\cite{Baird:2012cu}. As a consequence, widening the spin-priors leads to 
larger mass ratio distributions. Hence, different prior assumptions on mass ratio 
and component spins can lead to very different posterior distributions, and are
of key importance when interpreting the data.
In Appendix~\ref{app:PMTides}, from Eq.~\eqref{eq:delta_phase}, we see that this correlation may also reflect
on the estimate of $\tLam$ even in the case of high SNR signals, leading to an overall broadening
of the $\tLam$ posteriors.

\subsection{Tidal parameter and NS Radius}
\label{sec:inj_tides}
%=======================================

%=======================================
\begin{figure*}[t]
  \centering 
    \includegraphics[width=0.8\textwidth]{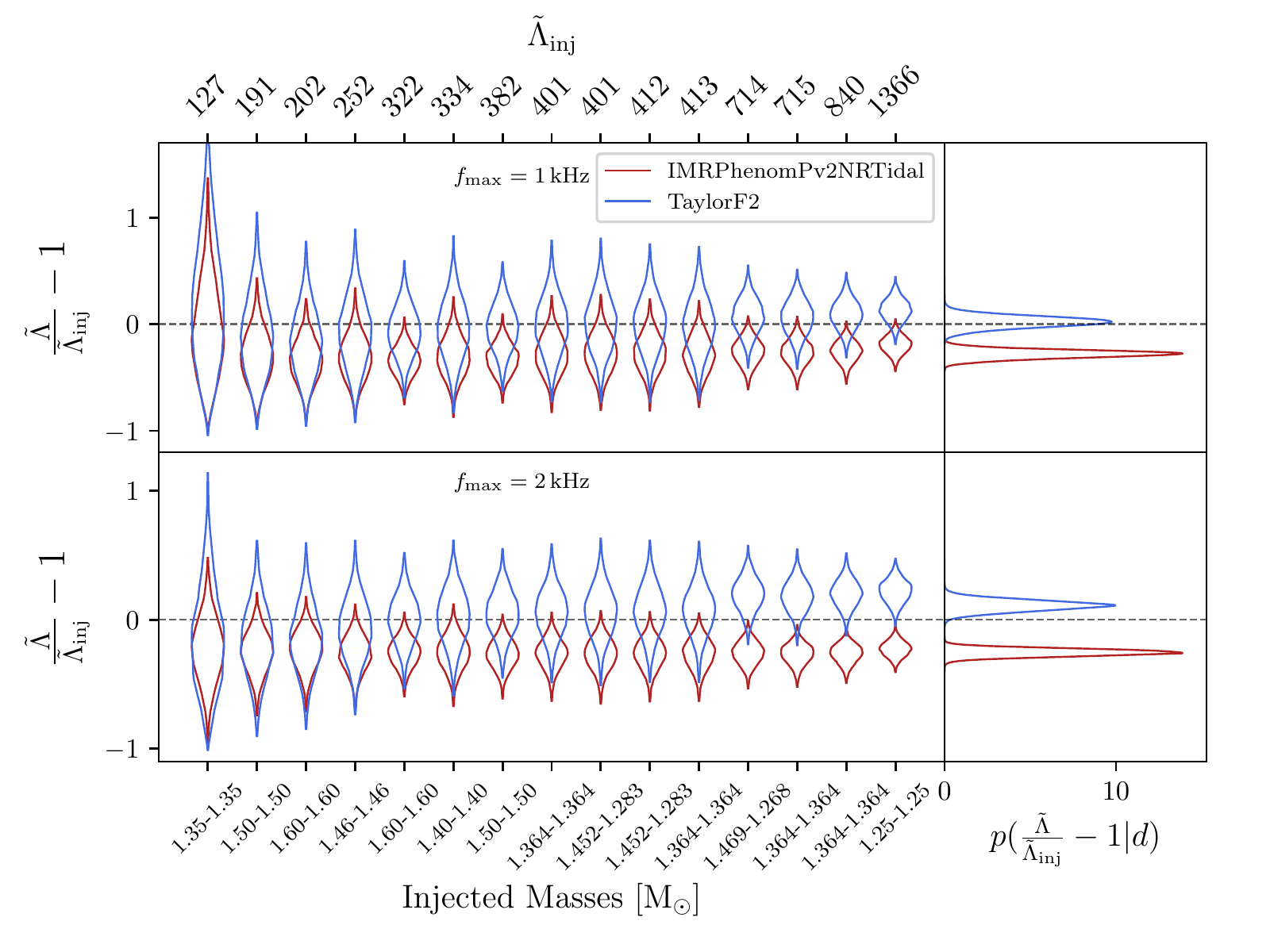}
    \caption{Main panel: violin plots of the fractional deviation between the injected values of $\tLam$ and the recovered posteriors.
    The color code depends on the approximant  employed for the PE (red for $\NRT1$, blue for $\TF2$), and the results 
    are displayed for  two different frequency cutoffs $f_{\rm max}$, $1$kHz (top) and $2$kHz (bottom). 
    As matter effects grow, the deviation between the two approximants and the {\tt TEOBResumS} baseline increases, 
    reaching approximately $\pm 20 \%$ when $\tLam^{\rm inj} = 1366$.
    On the right panel, we display the combined posterior distribution of the fractional deviation reweighted as described in the text. An increase of $f_{\rm max}$ negatively impacts the overall differences 
    in the recovery of $\tLam$ between models. This is especially true for $\TF2$: 
    PN waveform models are known to become less accurate close to merger frequencies.
    }
    
    \label{fig:lambda_violin}
\end{figure*}
%=========================

%==========================
\begin{figure*}[t]
  \centering 
    \includegraphics[width=\textwidth]{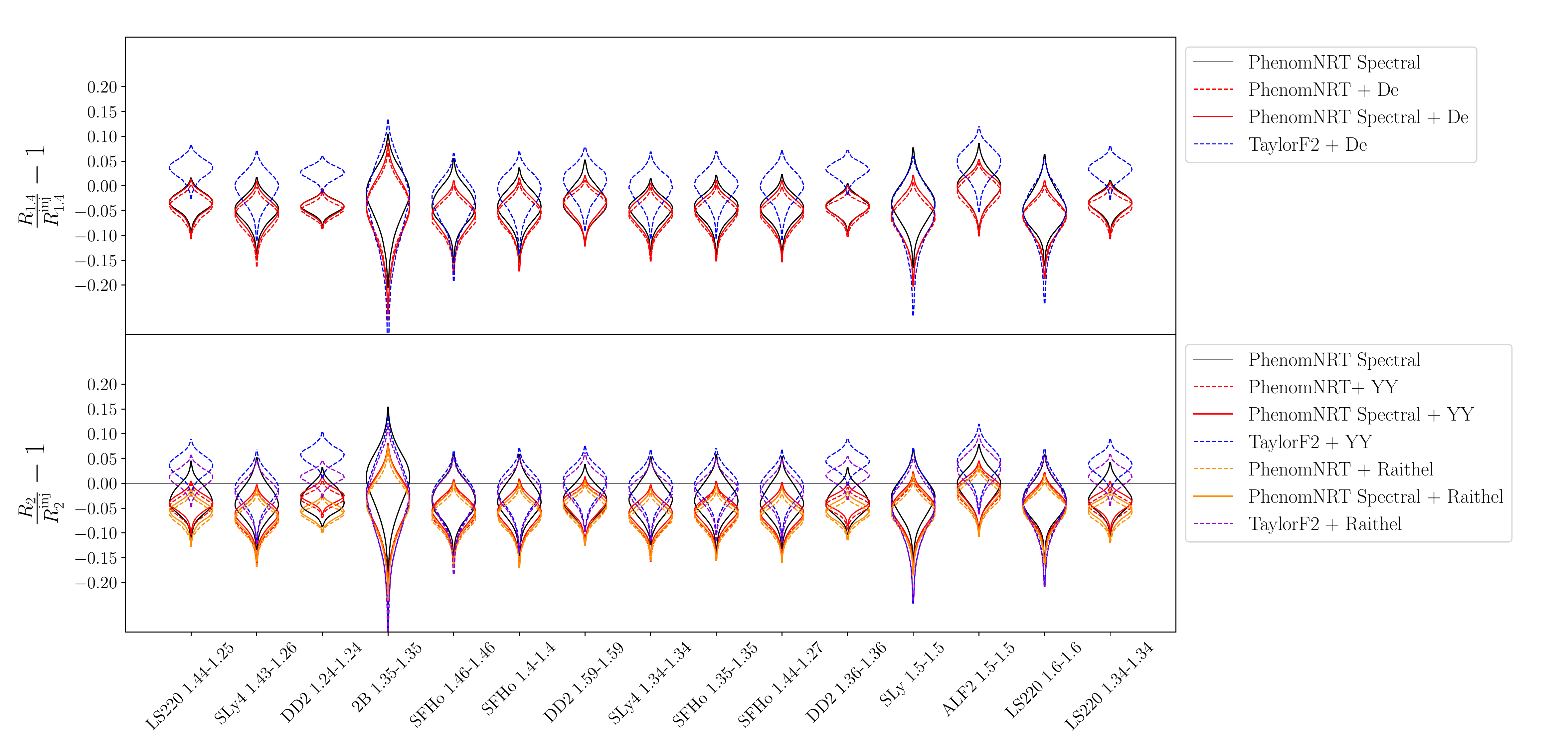}
    \caption{Fractional deviation between the real "injected"
    values $R_{\rm inj}$, computed for each signal listed in Table~\ref{tab:posteriors},
    and the distributions of $R$ computed either using the universal relations
    of Table~\ref{tab:UR} (colored lines) or by considering the 
    parameterized EOS posteriors and solving the NS structure equations (black lines).
    Whenever posteriors from the parameterized EOS ("spectral") runs are involved,
    we employ straight lines. Conversely, whenever we use posteriors from analyses performed by sampling
    $\Lambda_A$ and $\Lambda_B$ independently we use dashed lines.
    The under/over estimates displayed by $\NRT1$ and $\TF2$ in Figure \ref{fig:lambda_violin} 
    translate into similar biases on $R$, amounting up to $\pm 5\%$.}
  \label{fig:r_violin}
\end{figure*}
%================
% Table with notations
%================
 \begin{table}[t]
   \centering    
   \caption{ Summary of the the quasi-universal relations used in Sec.~\ref{sec:inj_tides}. While the De et al.~(De) 
   relation immediately links the radius of a $1.4 \Msun$ NS to the mass-weighted tidal deformability and chirp mass of a BNS system, 
   The Yagi-Yunes~(YY) and Raithel et al.~(R) relations 
   require the numerical inversion of $\tLam(R, q)$. For more detail, see Appendix~\ref{app:UR}}
   \begin{tabular}{c|c|c}        
     \hline
     Shorthands & References & Expressions \\
     \hline \hline
     De (De et al.) & \cite{De:2018uhw} & Eq.~\eqref{eq:De} \\
     YY (Yagi and Yunes) & \cite{Yagi:2015pkc, Yagi:2016bkt} & Eq.~\eqref{eq:BinaryLove}, \eqref{eq:CLambda} \\
     R (Raithel et. al) & \cite{Raithel:2019uzi} & Eq.~\eqref{eq:Raithel}\\
     \hline
   \end{tabular}
  \label{tab:UR}
 \end{table}

We now discuss systematics in the inference on tidal parameters and the effect on constraints on the NS radius.
Figure~\ref{fig:lambda_violin} shows the posterior distributions of the tidal 
parameters $\tLam$ recovered through PN and Phenomenological approximants. 
The values coming from the posterior samples are re-scaled by the {\it true} injected value, 
adopting the auxiliary parameter
\be
\varepsilon_{\tilde\Lambda}^{(i)} = \frac{\tilde \Lambda^{(i)}}{\tilde \Lambda_{\rm inj}^{(i)}} -1\ ,
\ee 
which encodes the fractional deviation from the injected value
for each simulated signal $i$.
We observe that, as the injected values of $\tLam$ increase, 
the relative uncertainties of the recovered posterior distributions decrease
 and modeling differences become more relevant 
(the median of the distributions are shifted with respect to zero). The combination of these 
two effects leads to evident biases in the recovered values. 
The overall  bias due to waveform effects 
is quantified by the combined posterior distribution $p(\varepsilon_{\tilde\Lambda}|d)$
shown in the right panel of Fig.~\ref{fig:lambda_violin}.
This quantity is estimated weighting each posterior distribution 
$p(\varepsilon_{\tilde\Lambda}^{(i)}|d)$
by the respective prior distribution $p(\varepsilon_{\tilde\Lambda}^{(i)})$,
computed from the prior distributions for $\tilde \Lambda^{(i)}$. 
The result is multiplied by the prior distribution $p(\varepsilon_{\tilde\Lambda})$ 
for the combined parameter $\varepsilon_{\tilde\Lambda}$,
taken as uniform in the range $[-2,+2]$, i.e.
\be
\label{eq:lambdat_prior}
p(\varepsilon_{\tilde\Lambda}|d) = p(\varepsilon_{\tilde\Lambda}) \prod_{i} \frac{p\Big(\varepsilon_{\tilde\Lambda}^{(i)}\Big|d\Big)}{p\Big(\varepsilon_{\tilde\Lambda}^{(i)}\Big)}\,,
\ee
where the index $i$ runs over all the injected binaries.
We find that {\tt IMRPhenomPv2NRTidal} systematically recovers lower values than those 
injected with {\tt TEOBResumS}, while {\tt TaylorF2} tends to systematically overestimate 
tidal parameters as matter effects increase, although it is able to capture 
the injected values for $\tLam \leq 400$.
These results can be understood in terms of the $Q_{\omega}$ analysis of
Sec.~\ref{sec:wf}, coupled to the relevant frequency ranges computed and 
discussed in Appendix~\ref{app:tidalinfo}. To summarize, 
the analyzed signals contain useful information
up to approximately 1kHz, depending on the source parameters.
We are then consistently in the situation where $f_{\rm thr}$ 
is larger than $f^{\tLam}_{5\%}$, whose values lie around
$240-300$ Hz.
Then, as shown in the third column of figure \ref{fig:Qomg}, 
for {\tt IMRPhenomPv2NRTidal} systematical differences in 
$\tLam$ are dominated by the tidal sector, which is more attractive 
than {\tt TEOBResumS} and leads to lower estimates of $\tLam$. 
The attractive point mass contribution of {\tt TaylorF2}, instead, leads to 
slight underestimates of the tidal parameters for values of $\tLam \approx 100$, 
while for $\tLam \approx 400$ it compensates the tidal sector.
The latter dominates for larger values of $\tLam$, and -- being too
repulsive -- causes overestimates of matter effects.

Translating information on the tidal parameters
of a NS into information on the NS EOS and radius
is not straightforward.
Given that waveform models do not explicitly depend on the NS radius, it is not possible to directly 
extract $R$ from GW data. It is necessary, instead, to rely on either some parameterization of the EOS~\cite{Read:2008iy, Raaijmakers:2018bln, Lindblom:2010bb, Lindblom:2013kra}, 
or on quasi-universal (EOS-insensitive) relations, which phenomenologically link macroscopic quantities of the binary between each others.
In particular, we employ the spectral parameterization of~\cite{Lindblom:2010bb, Lindblom:2013kra} and the EOS universal relations of De and Lattimer~\cite{De:2018uhw},
of Raithel et al~\cite{Raithel:2019uzi}, and of Yagi and Yunes~\cite{Yagi:2015pkc, Yagi:2016bkt}. The EOS-insensitve relations used here are
summarized in Appendix~\ref{app:UR}.

In the reminder of this subsection we focus on the implication of waveform systematics on the recovery 
of the NS radii and EOS reconstruction. We additionally gauge the further biases that can be introduced 
by employing quasi-universal relations for the recovery of $R$.
To do so, we apply the above UR 
to the analyses performed by sampling the component tidal 
parameters $\Lambda_i$ independently of each others, \textit{as well as} 
to (spectral) parameterized EOS runs.
Indeed, the parameterized posterior EOS obtained are usually employed 
in conjunction with the component mass posteriors $m_i$ to solve the 
TOV structure equations, and obtain a direct 
estimate of $R$. At the same time, however, given an EOS and the component masses, 
it is possible to compute $\tLam$, apply some UR and obtain another -- in
principle equivalent -- estimate of $R$.
This allows for a direct comparison of the effects of using universal relations
in place of parameterized EOS runs, independently of the choice of the sampling parameters 
(and, therefore, of the implied priors on $\tLam$).
Figure~\ref{fig:r_violin} shows the distributions of the
deviation in the estimates of $R_{1.4 \Msun}$ (top panel) and $R_2$ (bottom panel)
with respect to the real radii values corresponding to the parameters and EOSs listed in
Table~\ref{tab:posteriors}.
We find that {\tt IMRPhenomPv2NRtidal} tends to underestimate $R$, while {\tt TaylorF2}
behaves in the opposite way. 
The overall bias can amount up to approximately 
$\pm 5\%$ between {\tt TEOBResumS} and PN/phenomenological waveforms and $10 \%$ between $\NRT1$ and $\TF2$. 
Mirroring the behavior of $\tLam$, it becomes more relevant as tidal effects grow.
Additionally, all universal relations 
lead to slight underestimates of the values of R with respect to the ones recovered from spectral runs. 
We find that the true values of $R$ fall outside the $90\%$ credible intervals in a significant 
number of cases, especially when computing $R_{1.4 \Msun}$.
In our situation, with an injected EOB waveform,
we find that while this additional difference impacts negatively {\tt IMRPhenomPv2NRTidal} analyses, {\tt TaylorF2} runs would
 gain from using universal relations rather than a parameterized analysis.

\subsection{Faithfulness thresholds and PE biases}
\label{sec:PEbiases}
\begin{figure}[t]
  \centering 
    \includegraphics[width=0.5\textwidth]{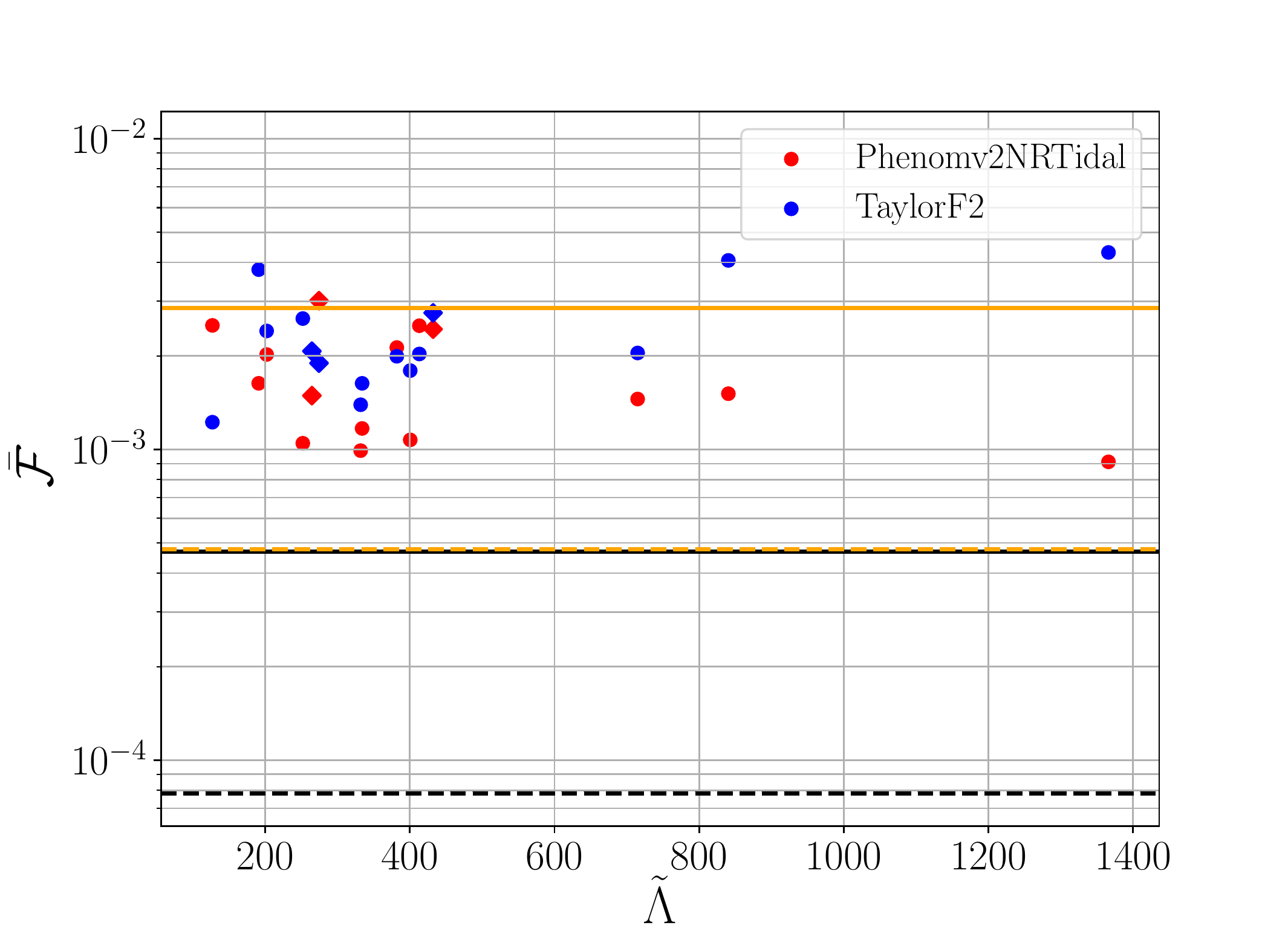}
    \caption{Unfaithfulness values between the {\tt IMRPhenomPv2NRTidal} (red) or 
      {\tt TaylorF2} (blue) waveforms and the injected {\tt TEOBResumS} waveforms.
      The plus and cross polarizations are projected on to the Livingston detector, and the source
      is fixed at the GW170817 sky location. Horizontal straight lines correspond to the treshold
      values of $\bar{\mathcal{F}}$ obtained with an SNR of $80$ (black) and $32.4$ (orange), computed through 
      Eq.~\ref{eq:LB} with $\epsilon = 6$ (straight) and $\epsilon=1$ (dashed). Dots correspond to waveforms with $q=1$, 
      while diamonds to signals with $q \neq 1$.}
 \label{fig:fbar_1khz}
\end{figure}

Finally, we apply the accuracy criteria of Sec.~\ref{sec:ana} to our data and show that, 
while criteria based on faithfulness alone are of little use to predict the presence of biases, 
an estimate of the parameter bias can be obtained using Eq.~\eqref{eq:bias_min}.
We begin by computing the unfaithfulness $\bar{\mathcal{F}}$ between waveform  
models evaluated with the same set of true parameters 
$\theta_\text{inj}$ through Eq.~\eqref{eq:faithfulness}. 
We place all sources in GW170817's sky location, and employ the
analytical {\tt aLIGODesignSensitivityP1200087} PSD~\cite{TheLIGOScientific:2014jea}, provided by {\tt pycbc}~\cite{pycbc}.
The results are summarized in Fig.~\ref{fig:fbar_1khz}.  
We find that both {\tt TaylorF2} and {\tt IMRPhenomPv2NRTidal} give values 
largely above the nominal threshold of $\bar{\mathcal{F}}=0.03$ (not shown in figure), 
which corresponds to ${\sim}10$\% of detection losses \cite{Cutler:1994ys, Lindblom:2008cm}.
When considering the thresholds provided by Eq.~\ref{eq:LB} in its weaker
formulation ($\epsilon^2=6$), we find that most signals fall above the 
value corresponding to GW170817's network SNR (straight orange line), and largely
below the threshold corresponding to a network ${\rm SNR}$ or $80$ (straight black line),
i.e the SNR above which all of our injections are
performed. By tightening the constraints and enforcing $\epsilon^2=1$ (dashed lines),
we find that already at the SNR of GW170817 none of the considered signals are 
faithful enough to ensure that no waveform systematics will be observed.
However, not all our injections give largely biased or
uncompatible results. These facts remark that 
these criteria give necessary but not sufficient rules to identify
biases and highlight the strong dependence of the criteria
themselves on the arbitrarily chosen value of $\epsilon^2$.

To obtain an estimate of the biased values $\tLamB$ we apply
Eq.~\eqref{eq:bias_min}, and minimize the quantity 
$ \sum_i (h^{\TEOB}_i - h^{\NRT1, \TF2}_i | h^{\TEOB}_i - h^{\NRT1, \TF2}_i)$,
where the sum is performed over the network interferometers considered (Livingston, Hanford
and Virgo, in our case).
In particular, for each waveform $h^{\NRT1, \TF2}$ we fix the intrinsic parameters
$(m_A, m_B, \chi_A, \chi_B)$ to their real injected values, and 
vary $\Lambda_A = \Lambda_B$ over the one-dimensional interval
$\tLam \in [\min(0, \tilde\Lambda_{\rm inj} - 500), \tilde\Lambda_{\rm inj} + 500]$. 
The simplifying choice of imposing $\Lambda_A = \Lambda_B$ can be justified by considering 
that in our injection study we were unable to distinguish $q = 1$ from
$q \neq 1$ systems. 
While this might not be true for more asymmetric systems than those studied in
the present paper, the issue can be easily circumvented by employing Binary-Love universal 
relations~\cite{Yagi:2015pkc, }. 
The straightforward procedure described leads to the values displayed in Table~\ref{tab:posteriors}.
We find that the $\tLamB$ values computed, while often
slightly overestimated with respect to the medians of the distributions of the
tidal parameters recovered through PE, fall into the $90\%$ $\tilde\Lambda$ credible
limits in the large majority of cases, thus providing a good approximation of the overall
behavior of the approximants employed.
Due to the overestimate of $\tLamB$, the bias 
$\Delta\tilde\Lambda_{B} = |\tLamB - \tilde\Lambda_{\rm inj}|$
is larger than the real bias 
$\Delta\tilde\Lambda_{\rm true} = |\tilde\Lambda^{\rm median} - \tilde\Lambda_{\rm inj}|$
for the {\tt TaylorF2} approximant, and smaller for {\tt IMRPhenomPv2NRTidal}.
Estimates of waveform systematics based on the above method might then be slightly optimistic (pessimistic)
when comparing {\tt TEOBResumS} to {\tt IMRPhenomPv2NRTidal} ($\TF2$).

\section{GW170817}
\label{sec:gw170817}
We now apply the approach developed and tested in the previous sections 
to the analysis of GW170817.

%====================

%====================
\begin{figure}[t]
   \centering
   \includegraphics[width=0.5\textwidth]{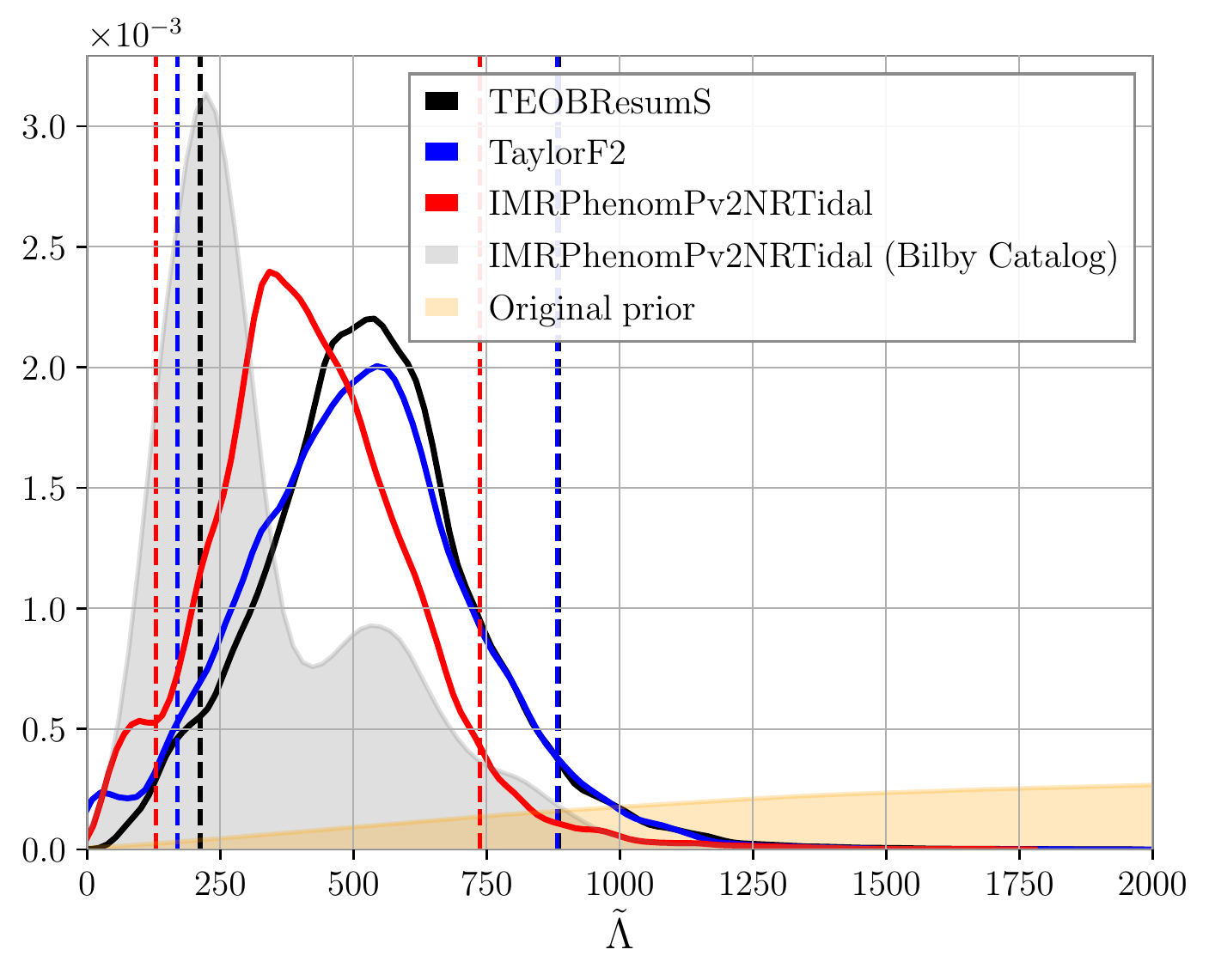}
   \caption{Analysis of GW170817 data. Marginalized one-dimensional $\tLam$ posteriors, obtained 
   analyzing the data up to $f_{\rm max}=1024$~Hz with three approximants: $\TF2$ (blue), $\NRT1$
   (red) and $\TEOB$ (black). The posteriors shown are reweighted to flat in $\tLam$ 
   prior, as is done in e.g \cite{LIGOScientific:2018mvr}. The public {\tt bilby} posteriors from 
   Ref.~\cite{Romero-Shaw:2020owr} (grey) are also displayed. Note that the {\tt bilby} analysis 
   uses $f_{\rm max}=2048$~Hz.}
   \label{fig:GW170817pos}
\end{figure}

\begin{figure}[t]
	\centering 
	\includegraphics[width=0.49\textwidth]{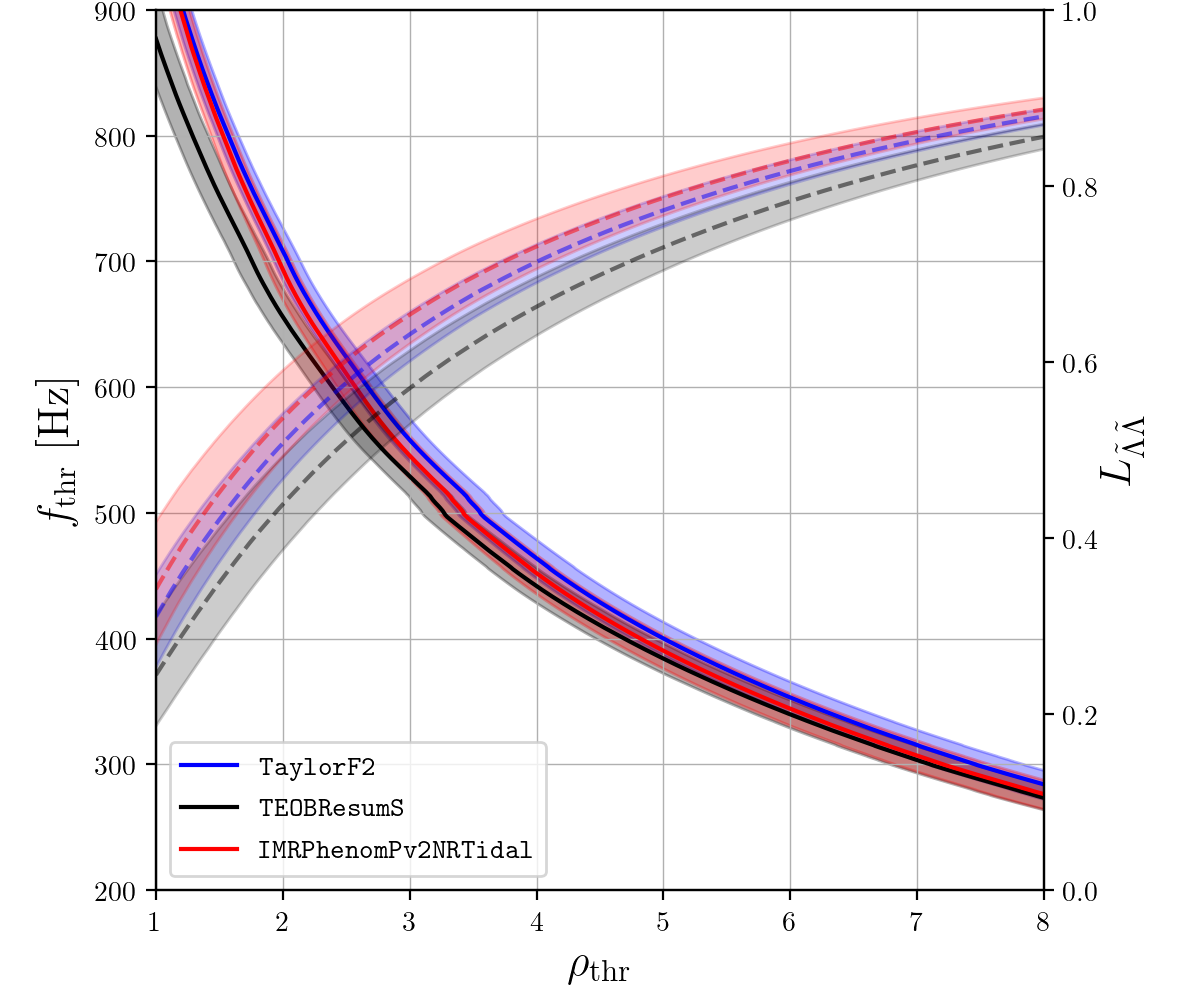}
	\caption{  Estimation of $f_{\rm thr}$ (solid lines) and $\Iloss$ (dashed lines) 
					from GW170817 posterior samples
					extracted with $\TF2$ (blue), $\NRT1$ (red) and $\TEOB$ (black). The shadowed bands represent the 90\% credible regions.
					In order to evaluate $f_{\rm thr}$, we use the identical waveform models involved in the extraction of the posterior samples and we set an cut-off frequency at merger $f_{\rm mrg}$, estimated 
					with the fits introduced in Ref.~\cite{Breschi:2019srl}.
					For the estimation of $\Iloss$, we limit ourselves to $6.5{\rm PN}$ phase corrections.
					The plotted values are computed using a network of three detectors (two LIGOs and Virgo) and setting $f_{\rm max}=2~{\rm kHz}$.
					For a single interferometer at $\rho_{\rm thr}=1$, we get $f_{\rm thr}\approx 700~{\rm Hz}$ for LIGO's detectors and $f_{\rm thr}\approx 250~{\rm Hz}$ for Virgo.}
	\label{fig:fM_170817}
\end{figure}

We perform a Bayesian analysis of GW170817 using the $\NRT1$, $\TF2$ and $\TEOB$
approximants, involving \texttt{pbilby} \cite{Smith:2019ucc}.
We adopt an almost identical configuration to the one presented in 
Ref.~\cite{Romero-Shaw:2020owr} (see also \cite{LIGOScientific:2018mvr}). 
In more detail, we consider a strain of 128 s around the GPS time $1187008882.43$~s. Data is
downloaded directly from the GWOSC \cite{Abbott:2019ebz}, in its cleaned and deglitched version (v2). We employ
the PSDs provided by \cite{LIGOScientific:2018mvr}, and fix the sky location to the one provided by EM constraints.
Further, as we are mainly interested estimating the intrinsic parameters of the source, we marginalize 
over distance, time and phase. The sampling is performed with uniform priors in chirp mass 
$\mathcal{M} \in [1.18, 1.21] \Msun$ and mass ratio $q \in [0.125, 1]$, with the additional constraints $m_A, m_B \in [1.001398, 4.313897948277728] \Msun$. The quadrupolar tidal coefficients $\Lambda_A, \Lambda_B$ are uniformly sampled in the interval $[0, 5000]$. The main differences w.r.t the analysis of 
Ref.~\cite{Romero-Shaw:2020owr} lie in (i) the different spin priors employed, which are taken to be 
aligned to the orbital angular momentum and such that $(\chi_A, \chi_B) \in [-0.05, 0.05]$,
and (ii) in the high-frequency cutoff of 1024 Hz that we impose (instead of the 2048~Hz of~\cite{Romero-Shaw:2020owr}).

Using the formalism of the Fisher matrix outlined in Sec.~\ref{sec:staterr},
we investigate in which frequency region the tidal information is effectively extracted,
according with the extracted posterior samples:
the Fisher's matrix element $I_{\tLam\tLam}$ 
has its main support in the frequency band from $200~{\rm Hz}$ to $1.5~{\rm kHz}$.
Subsequently, we compute $f_{\rm thr}$ according to Eq.~\eqref{eq:fM}
and, in order to achieve a more realistic result, 
we neglect the contributions above merger frequency $f_{\rm mrg}$, 
where this quantity is estimated using numerical relativity fits introduced in Ref.~\cite{Breschi:2019srl}.
As shown in Fig.~\ref{fig:fM_170817}, 
we find that the SNR of GW170817 is located at frequencies lower than ${\sim}700$ Hz 
(depending on the chosen value of $\rho_{\rm thr}$). More precisely, we can say that
the signal power enclosed above ${\sim}900~{\rm Hz}$ does not exceed an SNR of 1 (which roughly corresponds to 3\% of the total SNR), while the power above ${\sim}550~{\rm Hz}$ cannot contribute more than an SNR of 3 
(10\% of the total SNR). The large variability of $f_{\rm thr}$ with the
chosen value of $\rho_{\rm thr}$ indicates that a relatively small fraction of the SNR is accumulated over a rather large frequency interval.
The estimation of $f_{\rm thr}$ motivates our choice of $f_{\rm max}=1~{\rm kHz}$: indeed,
we do not expect to find a relevant portion of signal above this limit. Furthermore, this upper bound minimizes 
errors due to possible high-frequency noise fluctuations.
From the discussion of Sec.~\ref{sec:ana}, one expects the
masses of the binary to be measured rather
accurately. The reduced tidal parameter, instead, will be affected by significant 
statistical uncertainties: 
from the posterior samples, we estimate a loss of tidal information 
of $\Iloss \sim 20-50\%$, for $\rho_{\rm thr} \sim 1-3$.

The marginalized tidal parameter 
posteriors reweighted to flat in $\tLam$ prior are shown in Fig.~\ref{fig:GW170817pos}. 
All the measurements agree within 95\% confidence region, thus indicating that waveform
systematics are not the main source of uncertainty. However, the distributions
for the different approximants do suggest the presence of some systematic effects.
These posteriors should be interpreted in terms of the phasing
plots in Fig.~\ref{fig:Qomg} for $f\lesssim700\,$kHz (low frequency part of right panel) and $\tLam \lesssim1000$.
The phasing analysis of Sec.~\ref{sec:wf} shows that $\NRT1$ is more 
attractive than $\TEOB$ and $\TF2$; the systematic differences in the
relevant frequency regime are dominated by the tidal part ($\NRT1$ vs
others) or by a mixture of the point-mass and tides ($\TF2$ vs
$\TEOB$). This is consistent with the slightly smaller $\tLam$ measured with
the $\NRT1$ with respect to the other approximants and attributable to
the particular design of $\NRT1$ (PN tides at LO in the low frequency regime, $\TEOB$ in
middle regime frequency and NR data at higher frequencies; with LO
tides stronger than PN NLO, NNLO, and EOB tides at low frequencies, and
NR tides typically stronger than EOB tides~\cite{Bernuzzi:2012ci,Bernuzzi:2014owa}).
$\TEOB$ measurement is instead compatible with $\TF2$. This is again
understandable from the phasing plots discussed in Sec.~\ref{sec:wf}: for 
$\hat\omega\sim0.035-0.06$, the differences in the point-mass and tidal sector between the
approximants have opposite sign and partially compensate each other. 
Nonetheless, it is not possible from this analysis to identify whether 
a model is preferred by the data available, consistently with the 
conclusion of~\cite{TheLIGOScientific:2017qsa,Abbott:2018wiz}. We report in Table~\ref{tab:logB} 
the evidences given by the different approximants.
We conclude that systematics effects are observable in GW170817, 
but do not dominate the measurement of $\tLam$. These effects are 
nonetheless expected a priori from the phasing analysis of Sec.~\ref{sec:wf}. 

Note that our $\NRT1$ posteriors do not present the double-peak in $\tLam$ 
that is instead found in~\cite{TheLIGOScientific:2017qsa,Abbott:2018wiz}. 
The reason for this difference lies in the high-frequency cutoff imposed. 
This same effect had already been noticed in Ref.~\cite{Dai:2018dca}.
The authors, using the spin-aligned {\tt IMRPhenomDNRTidal} model~\cite{Husa:2015iqa,Khan:2015jqa} to analyze the data, together with
the relative binning technique~\cite{Zackay:2018qdy}, found a double-peak structure in the posterior 
of $\tLam$ with $f_{\rm max}=1.5$~kHz, that however disappeared when $f_{\rm max}$ 
was lowered to 1~kHz.
Repeating our analysis with $\NRT1$ and $f_{\rm max} = 2$~kHz, we too re-obtain the double peak in $\tLam$. 
The evidence of the newer analysis is, however, compatible to the one
reported in Table~\ref{tab:logB}:
$\ln p(d|\NRT1, 2 {\rm kHz}) = $ 521.860 $\pm$  0.103.
This implies that negligible SNR is accumulated above 1 kHz, and that the double peak is not to be interpreted 
as a physically motivated feature of the posteriors, but rather can be attributed to some high frequency noise fluctuation. This fact is also supported by the estimation of $f_{\rm thr}$.

Overall, we find consistent values for intrinsic parameters such as masses and spins with \cite{TheLIGOScientific:2017qsa,Abbott:2018wiz} 
and higher $\tLam$ values.
To translate the information on $\tLam$ to constraints on the NS radius $R$ we apply the UR of \cite{De:2018uhw} to the reweighted $\TEOB$ 
$\tLam$ posteriors and estimate the radius of a $1.4 \Msun$ NS.
We find $R_{1.4} = 12.5^{+1.1}_{-1.8}$ km. 
This value is slightly larger -- though still compatible -- than the one obtained in \cite{TheLIGOScientific:2017qsa}. 
The effect of the key choices of our analysis, i.e the high frequency cutoff employed, the use of $\TEOB$ and the low-spin priors 
imposed, is then that of pushing towards higher $R$ values and softer EOSs.
In the literature, additional radius estimates have been computed by including further astrophysical information.
We find our result, which focuses on the implications of GW data alone, to be in good agreement with the radii
obtained when additionally accounting for electromagnetic-priors
\cite{Radice:2018ozg} and the measurement given by NICER.
\cite{Raaijmakers:2019qny, Raaijmakers:2019dks}, which too favour $\tLam$ values larger than $\approx 200$.

%======================

%======================
 \begin{table}[t]
   \centering    
   \caption{Analysis of GW170817 data: log-evidences and corresponding standard deviations
   	computed using different waveform approximants: $\TEOB$, $\TF2$ and $\NRT1$. 
   The values obtained, according to standard Bayesian statistics, 
   indicate that it is not possible to identify a preferred waveform model exclusively 
   by relying on the GW170817 data.}
   \begin{tabular}{cc}        
     \hline
     Approximant & $\ln p(d|{\rm Approx.})$ \\
     \hline
     $\TF2 $ &  $523.078 \pm 0.102$ \\
     $\TEOB$ &  $522.585 \pm 0.102$ \\
     $\NRT1$ &  $522.261 \pm 0.103$ \\
     \hline
   \end{tabular}
  \label{tab:logB}
 \end{table}

\section{Tides inference with 3G detectors}
\label{sec:3G}

Third generation detectors such as Einstein Telescope \cite{Sathyaprakash:2011bh, Maggiore:2019uih} 
and Cosmic Explorer~\cite{Reitze:2019iox} are expected to start taking data in the late 2020s.
Their increased sensitivity at high frequencies will significantly
improve the detection of tidal signatures in the inspiral, and even
allow the detection of GWs from the remnant.
Typical SNRs expected for GW170817-like events detected by ET are
of the order of 1700. As a consequence, the importance of waveform systematics
is expected to further increase with respect to second generation detectors.

To summarize some the arguments of Sec.~\ref{sec:ana}, the SNR enters the determination of $\tLam$
through two main channels. Firstly, it determines the maximum useful frequency 
$f_{\rm thr}$ (see Eq.~\eqref{eq:fM}), above which variations of $\rho$ can be 
fully attributed to statistical fluctuations and which 
determines the regimes at which tidal measurements are performed. Secondly
it is related to the width of the distribution of the tidal parameter $\sigma_{\tLam} = \tLam^{95^{\rm th}\%ile} - \tLam^{5^{\rm th}\%ile}$.
If the signal is loud enough -- as is expected with ET and CE -- $f_{\rm thr}$ will be above merger frequency
for a large fraction of events. Therefore, when studying the signal with inspiral-merger only waveform models,
the effect of varying the SNR will mainly affect $\sigma_{\tilde\Lambda}$.
To obtain a quantitative estimate of $\sigma_{\tilde\Lambda}$ for 3G detectors 
we fit the values found in our injection study and extrapolate them to higher SNRs.
We find that a good approximation of the behavior of $\sigma_{\tilde\Lambda}$ over the SNR
range we considered is obtained by assuming that 
\be
\label{eq:sigmatL}
\sigma_{\tLam}(\rho) = \frac{c}{\rho - \rho_0} .
\ee
This functional form is valid only for $\rho > \rho_0$, in which case the denominator can be expanded 
as a geometrical series, and incorporates the corrections to the leading
order $1/\rho$ asymptotical behavior expected from the Fisher Matrix analysis.
Fitting Eq.\eqref{eq:sigmatL} to the data we find $(c, \rho_0) = (7497.965751, 63.092154)$ for {\tt TaylorF2} and $(4372.214662, 66.778801)$ for {\tt IMRPhenomPv2}.
As could already be observed from Fig.~\ref{fig:lambda_violin}, $\NRT1$
constrains the tidal parameter better than its PN counterpart: $\sigma_{\tLam}^{\rm Phenom}(\rho)$ is (almost) parallel
to $\sigma_{\tLam}^{\rm TaylorF2}(\rho)$ but shifted to lower values.
To obtain a unique estimate of $\sigma_{\tLam}$ we compute the mean value 
$\bar{\sigma}_{\tLam} = (\sigma_{\tLam}^{\rm Phenom} + \sigma_{\tLam}^{\rm TaylorF2})/2$.

\begin{figure*}[t]
   \centering
   \includegraphics[width=0.49\textwidth]{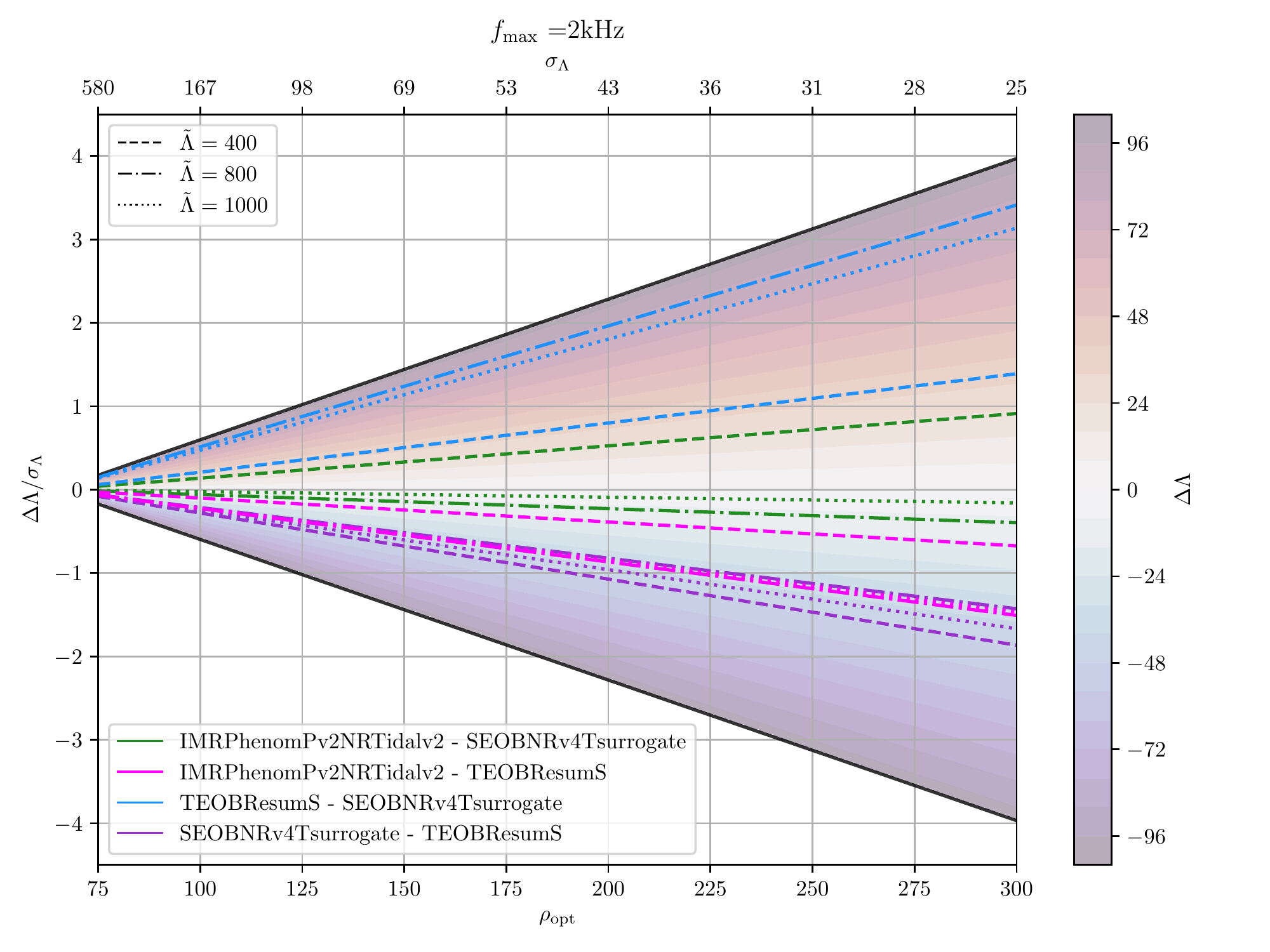}
   \includegraphics[width=0.49\textwidth]{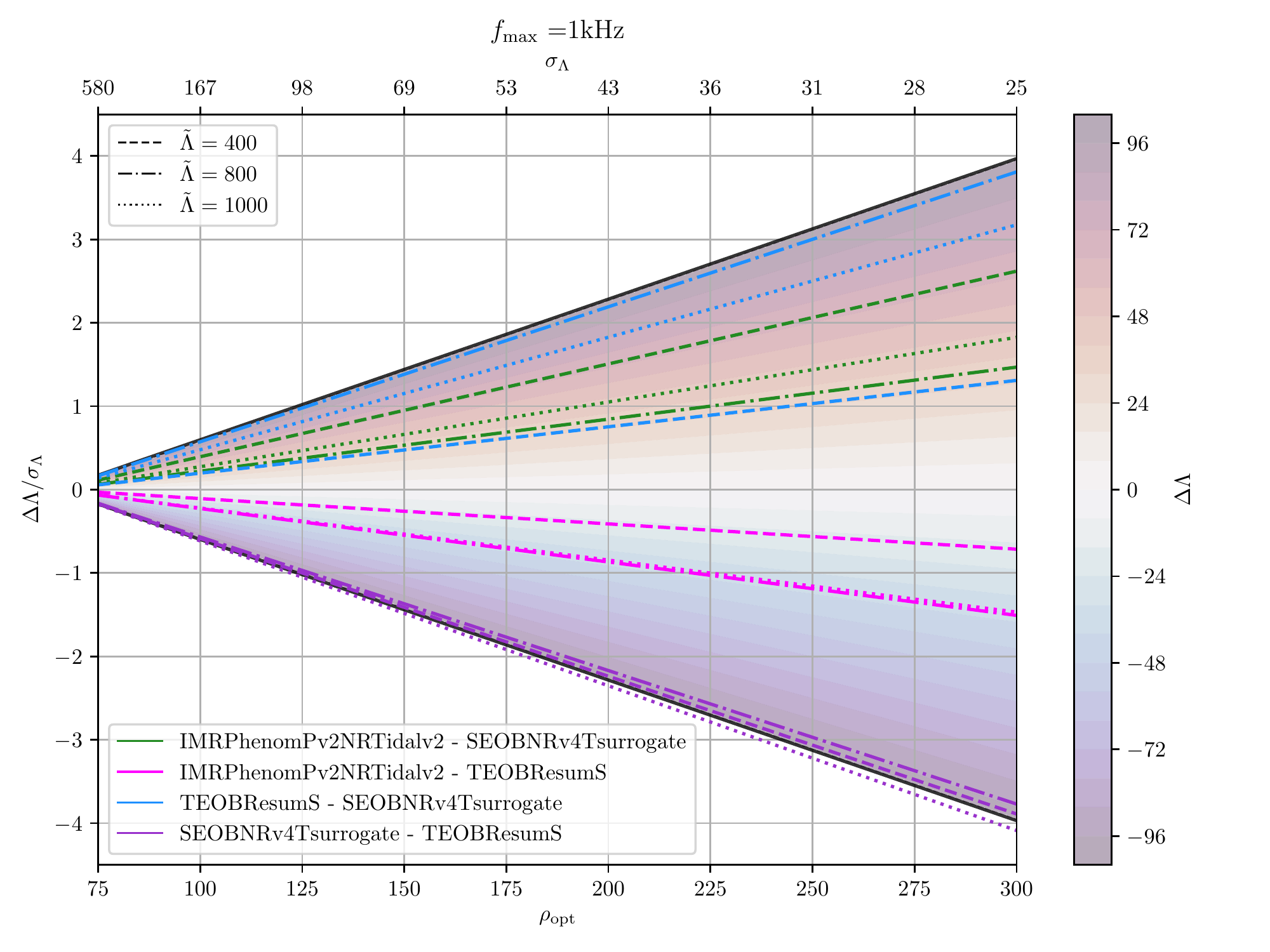}
     \caption{The ratio between systematic effects $\Delta\tLam$ and statistical uncertainties $\sigma_{\tLam}$, shown
     as a function of the SNR $\rho$ for a range of different $\Delta\tLam\in [-100, +100]$.
     Colored lines refer to values of $\Delta\tLam/\sigma_{\tLam}$ estimated between a baseline approximant 
     Y and a recovery approximant X, labelled as ``X-Y'' in the bottom-left legend, and computed for two different frequency
     cutoffs (left and right panels) and three different values of $\tLam$ (dashed, dotted and dash-dotted lines).
     We find that $\Delta\tLam/\sigma_{\tLam} \approx 1$ at SNRs ranging from 175-200 for all $\tLam$ values. Therefore, with
     3G detectors, all
     the current approximants will appear to be statistically inconsistent.
        \label{fig:bias3G}
      }
\end{figure*}

The expression of $\bar{\sigma}_{\tLam}$ can then be used to compute the SNR
at which two independent measurements $\tLam_1$ and $\tLam_2$, whose difference
we denote as $\Delta\tLam$, become statistically inconsistent.
Figure~\ref{fig:bias3G} shows the quantity $\Delta\tLam/\bar{\sigma}_{\tLam}$ as a function of the 
optimal SNR $\rho$ for values of $\Delta\tLam \in [-100, 100]$.
When $\Delta\Lambda/\bar{\sigma}_{\tLam} \approx 1$, statistical fluctuations
are of the same order of magnitude as systematical effects. For $|\Delta\tLam| \approx 100$, we see that this
condition is satisfied already at the threshold $\rho \approx 125$. As $|\Delta\tLam|$ decreases, 
the threshold SNR increases, reaching $\rho \approx 300$ in correspondence of a $\Delta\tLam \approx 20$.

The above considerations are independent of the exact waveform models employed, and
do not tackle the issue of estimating the $\Delta\tLam$ associated to two specific 
chosen approximants. While it is clear from the injection study of Sec.~\ref{sec:analysis}
that large $\Delta\tLam$ are to be expected when employing $\TF2$ and $\NRT1$, we take a
step further and qualitatively estimate the bias $\Delta\tilde\Lambda = \tLamB - \tLam$ 
through Eq.~\eqref{eq:bias_min} for two additional state-of-the-art approximants, 
{\tt IMRPhenomPv2NRTidalv2}~\cite{Dietrich:2019kaq} 
and {\tt SEOBNRv4Tsurrogate}~\cite{Lackey:2018zvw}.
We thus compare the latter and 
{\tt TEOBResumS} in pairs and report the differences with respect to
two baselines ({\tt TEOBResumS} and {\tt SEOBNRv4Tsurrogate}).
Following the procedure described in Sec.~\ref{sec:PEbiases} 
we consider values of $\tLam$ equal to $400$, $800$ and $1000$, 
place the sources in GW170817's location
and employ the {\tt EinsteinTelescopeP1600143} PSD~\cite{Evans:2016mbw}. 
We compute waveforms from $30$ to $2048$ Hz (left panel) or $1024$ Hz (right panel).
Results are again displayed in Fig.~\ref{fig:bias3G}.

We find that both {\tt SEOBNRv4Tsurrogate} 
and {\tt IMRPhenomPv2NRTidalv2} ``underestimate''
the values of $\tLam$ of the {\tt TEOBResumS} baseline (right panel),
and that the $|\Delta\Lambda|$ found are  
always below $\approx100$. This indicates that tides are stronger 
in the  {\tt SEOBNRv4Tsurrogate} and {\tt IMRPhenomPv2NRTidalv2} 
models than in {\tt TEOBResumS}.
When restricting below 1kHz (large $\tLam$) the systematic bias in
$\tLam$ due to the differences between 
{\tt IMRPhenomPv2NRTidalV2} and {\tt TEOBResumS} is
${\lesssim}2\sigma$ corresponding to $\Delta\tLam\pm50$, while it
varies ${\sim}2-4\sigma$ when considering differences with respect to
{\tt SEOBNRv4Tsurrogate}. This indicates that the differences between
{\tt IMRPhenomPv2NRTidalv2} and {\tt TEOBResumS} are 
mostly related to the modeling of tides at high-frequencies, while the
tides in the EOB models differ from each other already at lower frequencies.

Some caution is needed when interpreting the results
obtained for the different waveform approximants: in Sec.~\ref{sec:PEbiases} we have seen 
that at times the estimated $\tLamB$ would overestimate $\tLam^{\rm median}$ by up to $100$.
This difference was acceptable at the injected SNRs, but indicates that our estimate
might not precise enough at the SNRs which characterize 3G detectors.
Nonetheless, we expect the behavior of the approximants (i.e, their being more/less attractive)
to be correctly captured.

Overall, our findings indicate that above SNR $\approx 100-200$ $\sigma_{\tLam}$ will be small 
enough that the models will appear to be fully inconsistent between
each others. The estimated systematic biases reflect differences in
the tidal modeling at frequencies corresponding to the very last
orbits and thus accessible to NR. We stress that at frequencies
$\hat\omega_{22}\gtrsim0.06$ the NS are in contact and the waveform
modeling based on tidal interactions can only be considered an
effective description, since the dynamics is dominated by
hydrodynamics~\cite{Bernuzzi:2012ci}. We demonstrate in
Appendix~\ref{app:NRacc} that current NR simulations are 
not sufficiently accurate to produce faithful waveforms. New, more 
precise NR simulations appear crucial to further develop tidal 
waveform models for future detectors.
%===================

%===================

\section{Conclusions}
\label{sec:conc}

In this paper we discussed a possible approach for the analysis of waveform systematics in the estimation of tidal effects in BNS.
We demonstrated the effectiveness of our method in a mock experiment using a large set of injected signals and applied the method to GW170817.
We recommend to use this method for future analysis and point out that
the approximants used for the main analysis of GW170817 should be
significantly improved for future robust analysis at SNR ${\sim}80$
and beyond.
We expand on these conclusions here below.

The bottom-up method employed in this work is composed
of three steps. First, the waveform approximants should be compared
using the $Q_\omega$ analysis in order to understand the effect of the
modeling choices (and the physics implemented in the models) on the GW
phase. The $Q_\omega$ diagnostic is key to determine the waveforms'
differences and it is free from the ambiguities introduced in the
phase comparisons by the time/phase shift. 
Second, it is important to identify what is the frequency regime at
which the tidal information is effectively extracted. This can be
accomplished by computing fractional losses $L_{ii}$ defined in
Sec.~\ref{sec:ana}. 
Third, the PE results should be interpreted in terms of the $\Delta
Q_{\omega}$ analysis on the relevant frequency interval.  

Our mock experiments show that this procedure is effective in
identifying the main baiases introduced by the waveform models. Note
in this respect that the ``target'' model used for the computation
of the $\Delta Q_{\omega}$ should be chosen amongst those that are
considered sufficiently faithful on the 
relevant frequency regime. For example, for analyzing biases at low
frequencies, the target model should contain maximal analytical
information (vs minimal fitting), high-order Taylor or EOB models represent the best
choice in this respect. At very high frequencies, numerical
relativity data would be the best choice, although the accuracy of
the data is not yet sufficient for robust statements
(See Appendix~\ref{app:NRacc}).

The analysis of GW170817 shows that the measurement of the tides is
essentially free of systematic effects if performed up to $1\,$kHz \cite{Radice:2018ozg, Dai:2018dca}.
Extending the analysis to higher frequencies introduces some waveform
effects, albeit still compatible with others in the 90\% confidence
region. In particular, comparing our results to Fig.~9
of~\cite{LIGOScientific:2018mvr}, we observe a shift in the posterior
of $\tLam$ computed with $\NRT1$ that can be fully understood from
the $Q_\omega$ phasing analysis presented here. When applying
$\NRT1$ to $2\,$kHz, the posteriors have a double peak that is not
present with the $1\,$kHz cutoff.
The similar inferences using the $\TF2{}$ and $\TEOB$ approximants are instead related to the fact that, in the relevant
frequency regime, the differences between the $\TF2{}$ point-mass and the
tidal have opposite signs and partly compensate each other (see
Fig.~\ref{fig:Qomg}).

Applying the UR of \cite{De:2018uhw} to the $\tLam$ values obtained in our GW170817
re-analysis we obtain a new measurement of $R_{1.4 \Msun}$ which -- based exclusively
on the information gathered from GW data -- is in good agreement with results
coming from independent astrophysical observations, i.e the NICER radius measurement 
and the information coming from EM observations \cite{Radice:2018ozg, Raaijmakers:2019qny}.

Significant waveform systematics are to be
expected for GW170817-like signals already for the current advanced detectors at 
design sensitivity. Note these high-SNR signals are the only/best candidates
for an actual {\it measure} (vs. upper limit) of the tidal parameters and EOS constraints.
At design sensitivity, the expected bias in the reduced tidal
parameter using $\TF2$ and $\NRT1$ is about $2-\sigma$ (for
average BNS parameters as quantified in Fig.~\ref{fig:lambda_violin}).
This would reflect in systematics on the NS radius of about
$1$~km (10\%), that are comparable or well above of the
current best estimates of the NS radius, also including
electromagnetic constraints \cite{Annala:2017llu, Abbott:2018exr, De:2018uhw, Radice:2018ozg, Capano:2019eae}. 

Moving to higher sensitivities and 3G detectors, 
we estimate that the systematics between the approximants that currently
have the smallest differences among themselves 
become dominant over statistical errors at SNR 200
and for $\tLam\gtrsim400$ (Fig.~\ref{fig:bias3G}). 
This implies that EOS constraints from the potentially 
most informative (and rare) events will be harmed by 
tidal waveform systematics.

\begin{acknowledgments}
  We thank Jocelyn Read, Derek Davis and Katerina Chatziioannou
  for useful discussions and comments on the manuscript.
  R.~G. acknowledges support from the Deutsche Forschungsgemeinschaft
  (DFG) under Grant No. 406116891 within the Research Training Group
  RTG 2522/1. 
  M.~B. and S.~B. acknowledge support by the EU H2020 under ERC Starting
  Grant, no.~BinGraSp-714626.  
  M.~B.~ acknowledges support from the Deutsche Forschungsgemeinschaft
  (DFG) under Grant No. 406116891 within the Research Training Group
  RTG 2522/1. 
  Data analysis was performed on the supercomputers ARA in Jena and
  ARCCA in Cardiff. We acknowledge the computational resources provided
  by Friedrich Schiller University Jena, supported in part by DFG grants INST 275/334-1 FUGG and INST 275/363-1
  FUGG, and Cardiff University, funded by STFC grant ST/I006285/1. 
  Data postprocessing was performed on the Virgo ``Tullio'' server 
  in Torino, supported by INFN.
  This research has made use of data obtained 
  from the Gravitational Wave Open Science Center (https://www.gw-openscience.org), 
  a service of LIGO Laboratory, the LIGO Scientific Collaboration and the 
  Virgo Collaboration. LIGO is funded by the U.S. National Science Foundation. 
  Virgo is funded by the French Centre National de Recherche Scientifique (CNRS), 
  the Italian Istituto Nazionale della Fisica Nucleare (INFN) and the 
  Dutch Nikhef, with contributions by Polish and Hungarian institutes.
\end{acknowledgments}

\appendix

\section{Effect of the point mass sector on $\tLam$ }
\label{app:PMTides}
In this apendix, we explicitly show how uncertainties in the point mass phase
(of both statistic and systematic nature)
can affect the determination of the tidal parameter $\tLam$.
Starting from Eq.~\eqref{eq:rho2}, writing
 $\exp(i\Delta\Psi) = \cos(\Delta\Psi) + i \sin(\Delta\Psi)$,
 and expanding the cosine around $\Delta\Psi \approx 0$ the SNR becomes
\be
\begin{split}
\rho &= 
\frac{4}{\sqrt{(h|h)}} \int \frac{\tilde{A}_d \tilde{A}_h}{S_n} \left[ 1 -\frac{(\Delta\Psi)^2}{2} + O\left(\Delta\Psi^3\right)\right] {\rm d}f \\
&\simeq \rho_{\rm opt} - \frac{2}{\sqrt{(h|h)}}\int \frac{\tilde{A}^2 (\Delta\Psi)^2}{S_n} {\rm d}f\,, \label{eq:I}
\end{split}
\ee
where the last step assumes $\tilde{A}_h \approx \tilde{A}_d = \tilde{A}$.
By defining $\bar{\params}$ as
the set of parameters such that $\Delta\Psi (\bar{\params}; f) \approx 0$ over the whole frequency range considered $[f_{\rm min}, f_{\rm max}]$,
and expanding $\Delta\Psi$ in Eq.~\eqref{eq:I} around $\bar{\params}$,
the second integral in Eq.~\eqref{eq:I} can be connected to the Fisher matrix 
\be
\label{eq:int}
\int \frac{\tilde{A}^2 (\Delta\Psi)^2}{S_n} {\rm d}f \approx \int \frac{\tilde{A}^2 \partial_i\Psi_h \partial_j\Psi_h}{S_n} \Delta \theta^i \Delta \theta^j {\rm d}f
\ee
with $\partial_i = \partial/\partial\theta^i$ and $\Delta\theta^i =
\theta^i - \bar{\theta}^i$ (repeated indeces imply a summation).
Under the assumption of high SNR, the integrals over $f$ in
Eq.~\eqref{eq:int} can be split as
\be
\label{eq:regimes}
- \frac{\Delta\theta^l \Delta\theta^m}{2}\int_{f_{\rm min}}^{f_c} I_{l,m}{\rm d}f  - \frac{\Delta\theta^{\tLam} \Delta \theta^{\tLam}}{2}\int_{f_c}^{f_{\rm max}} I_{\tLam, \tLam}{\rm d}f,
\ee
where $f_c$ is a ``cutoff frequency'' that identifies the beginning of the relevant frequency support of $I_{\tLam, \tLam}$ (see Eq.~\eqref{eq:fLamMinMax})
and has the value of $\approx 300\,$Hz for fiducial BNS.
Eq.~\eqref{eq:regimes} clearly shows the different frequency regimes at which the parameters are measured during PE. $\bar{\M}, \bar{q}$ and $ \bar{\chi}$
are determined during the early inspiral $(f \leq f_c)$; $\bar{\tLam}$ at higher frequencies $(f \geq f_c)$. 
Sampling methods will tend to recover the parameters $\params \rightarrow \bar{\params}$. However, due to the varying sensitivity of the detector
over different frequency ranges, the parameters measured during the early inspiral $\params_{\rm insp} = ({\M, q, \chi})$ converge faster 
than tidal parameters. Let's then go back to Eq.~\eqref{eq:int} and express its left hand side as
\be
- \frac{\Delta\theta^l \Delta\theta^m}{2}\int_{f_{\rm min}}^{f_c} I_{l,m}{\rm d}f - \frac{1}{2}\int_{f_c}^{f_{\rm max}} \frac{\tilde{A}^2 (\Delta\Psi)^2}{S_n} {\rm d}f.
\ee
The first integral has, again, been expanded about the set of parameters $\bar{\theta}_\text{insp}$. Taking the limit $\params_{\rm insp} \rightarrow \bar{\params}_{\rm insp}$, its contribution tends to zero by definition. The remaining second integral can be explicitly written as:
\be
\label{eq:delta_phase}
- \frac{1}{2}\int_{f_c}^{f_{\rm max}} \frac{\tilde{A}^2}{S_n} \left[\Delta\Psi^\text{PM}(\bar{\theta}_\text{insp}, \tLam=0) + \Delta\Psi^{T}(\bar{\theta}_\text{insp}, \tLam)\right]^2 \,{\rm d}f,
\ee
where we have separated $\Delta\Psi$ into its point mass ($\Delta\Psi^{\rm PM}$) and tidal ($\Delta\Psi^{ T}$) contributions.
Critically, $\Delta\Psi^\text{PM}(\bar{\params}_{\rm insp},
\tLam=0)$ is not necessarily close to zero above $f_c$, as
the parameters $\bar{\theta}_{\rm insp}$ are determined over
a different regime, and chosen to minimize
$\Delta\Psi^\text{PM}(\bar{\theta}_{\rm insp}, \tLam=0)$ below $f_c$.
The value $\bar{\tLam}$ therefore will have to minimize not only $\Delta\Psi^T$ over $[f_c, f_{\rm max}]$, but rather the sum of $\Delta\Psi^T$ and $\Delta\Psi^\text{PM}$. 
This means that both the tidal and the point mass sectors of a waveform model can introduce biases in the recovery of tidal parameters, and that overall phase differences accumulated over $f_c$ are absorbed mainly by $\tilde\Lambda$.

\section{Tidal information}
\label{app:tidalinfo}

In this appendix, we apply the method presented in Sec.~\ref{sec:staterr}
to the the signals involved in the PE studies of Sec.~\ref{sec:analysis},
proving that the injections are actually performed in an informative framework
for the tidal parameter, in which statistical fluctuations cannot be considered as 
the dominant source of the biases observed in the tidal parameter (see Fig.~\ref{fig:lambda_violin}).

Tab.~\ref{tab:infoinject} shows the values of the 
frequency support $[f_{5\%}^{\tLam},f_{95\%}^{\tLam}]$
defined in Eq.~\eqref{eq:fLamMinMax} computed for the injected signals,
including all the detectors involved in the analysis.
For all the cases,$f_{\rm mrg} > 1~{\rm kHz}$, indicating the presence of signal in the high-frequency regime,
and $f_{95\%}^{\tLam} > 1~{\rm kHz}$, meaning that the tidal contributions are relevant above this value.
Furthermore, Tab.~\ref{tab:infoinject} reports the values of $f_{\rm thr}$ and $\Iloss$,
defined respectively in Eq.~\ref{eq:fM} and Eq.~\ref{eq:Iloss},
computed for the same signals for $\rho_{\rm thr}=1,3$.
For $\rho_{\rm thr} =1 $, we have $f_{\rm thr} > 1~{\rm kHz}$, showing that the signal power is 
relevant above this threshold. For this values, $\Iloss \le 30\%$.
These facts are reflected in a lower variance on the posterior distribution for $\tLam$
coming from the PE analyses with $f_{\rm max}=2048~{\rm Hz}$ with respect to
the ones with $f_{\rm max}=1024~{\rm Hz}$.
Finally, for all the injected signals, we have $f_{\rm thr}> f^{\tLam}_{5\%}$, which
proves that these data contains information on the tidal parameter in an accessible frequency range.

\begin{table}[h]
	\caption{Values of $f^{\tLam}_{5\%}$, $f^{\tLam}_{95\%}$,
		$f_{\rm thr}$ and $\Iloss$ computed for the 
		signals involved in the injection studies, Sec.~\ref{sec:analysis}. 
		We recall that the injected signals have extrinsic properties identical to the maximum-posterior 
		parameters of GW170817~\cite{TheLIGOScientific:2017qsa}. 
		The reported values are estimated with a three detector network (two LIGOs and Virgo)
		at design sensitivity using ${\tt TEOBResumS}$ waveform model.}
	\label{tab:infoinject}
	\resizebox{0.49\textwidth}{!}{\begin{tabular}{cccc|cc|cc|cc}
			\hline
			\hline
			EOS   & $M$         & $q$ &$f_{\rm mrg}$ & $f_{5\%}^{\tLam}$ & $f_{95\%}^{\tLam}$ & \multicolumn{2}{c|}{$\rho_{\rm thr} = 1$} & \multicolumn{2}{c}{$\rho_{\rm thr} = 3$} \\
			& {[}$\Mo${]} &                      &     {[}Hz{]}                & {[}Hz{]}              & {[}Hz{]}              & $f_{\rm thr}$ {[}Hz{]}         & $\Iloss$                  & $f_{\rm thr}$ {[}Hz{]}         & $\Iloss$        \\ 
			\hline
			\hline
			DD2   & 2.71        & 1.00        & 1287  & 245   & 1460                  & 1085                   & 0.18                          & 731                    & 0.52            \\
			LS220 & 2.68        & 1.00    & 1366    & 259        & 1800                  & 1152                   & 0.23                         & 740                    & 0.57            \\
			LS220 & 2.69        & 0.86    & 1241  & 242        & 1332                  & 1055                   & 0.15                    & 731                    & 0.51            \\
			SFHo  & 2.71        & 1.00      & 1426 & 271        & 1825                  & 1207                   & 0.23                       & 766                   & 0.59            \\
			SFHo  & 2.72        & 0.88         & 1416   & 278    & 1862                  & 1252                   & 0.25                 & 772                   & 0.61            \\
			SLy   & 2.68        & 1.00        & 1588     & 273  & 1746                  & 1211                   & 0.22                       & 772                   & 0.60            \\
			SLy   & 2.69        & 0.88     & 1480      & 272             & 1816                  & 1208                   & 0.23                       & 766                   & 0.59            \\
			DD2   & 2.48        & 1.00   & 1206   & 240                  & 1666                  & 1033                   & 0.21                       & 693                    & 0.55            \\
			DD2   & 3.18        & 1.00         & 1192   & 249                   & 1715                  & 1125                   & 0.19                         & 782                   & 0.49            \\
			2B    & 2.70        & 1.00          & 1646  & 293                   & 1834                  & 1311                   & 0.24                        & 804                   & 0.63            \\
			SLy   & 3.00        & 1.00         & 1540   & 278                   & 1744                  & 1254                   & 0.20                    & 828                   & 0.56            \\
			LS220 & 3.20        & 1.00       & 1288         & 255                   & 1443                  & 1332                   & 0.30                    & 826                   & 0.63            \\
			SFHo  & 2.92        & 1.00     & 1449    & 281                   & 1874                  & 1285                   & 0.24                         & 802                   & 0.59            \\
			SFHo  & 2.80        & 1.00     & 1519    & 273                   & 1698                  & 1222                   & 0.20                & 788                   & 0.58            \\
			ALF2  & 3.00        & 1.00      & 1299       & 250                   & 1395                  & 1121                   & 0.15                      & 787                   & 0.50           \\
			\hline
			\hline
	\end{tabular}}
\end{table}

\section{Faithfulness of Numerical Relativity waveforms}
\label{app:NRacc}

\begin{table}[t]
\caption{Faithfulness values $\mathcal{F}$ computed considering
  frequencies from $f_{\rm low}$ to $f_{\rm mrg}$ between simulations
  with the same intrinsic parameters and two different resolutions,
  extracted at $r/M = 1000$. The source is situated in the same sky
  location as GW170817, and the waveform polarizations $h_+$ and
  $h_\times$ are computed and projected on the Livingston detector. We
  employ the {\tt aLIGODesignSensitivityP1200087}~\cite{TheLIGOScientific:2014jea} 
  PSD from {\tt pycbc}
  \cite{pycbc} to compute the matches, and compare the values obtained to
  the thresholds $\mathcal{F}_{\rm thr}$ calculated with
  Eq.~\ref{eq:LB} with $\epsilon^2=1$ or $\epsilon^2=N$. A
  tick~\cmark~indicates that $\mathcal{F} > \mathcal{F}_{\rm
    thr}$. Conversely, a cross~\xmark~indicates that $\mathcal{F} <
  \mathcal{F}_{\rm thr}$.} 
\label{tab:nr_acc}
\begin{tabular}{c|c|c|cc|cc|cc}
\hline  \hline
Sim & n\footnote{Number of grid point (linear resolution) of the
  finest grid refinement,
  roughly covering the diameter of one NS} & $\mathcal{F}$ & \multicolumn{6}{c}{SNR}\\
& & &  \multicolumn{2}{c}{14} & \multicolumn{2}{c}{30} &\multicolumn{2}{c}{80} \\
& & & $N=6$ & $1$ & $N=6$ & $1$ & $N=6$ & $1$\\   
\hline                
{\tt BAM:0011} & [96, 64]   & 0.991298 & \cmark & \xmark & \xmark & \xmark & \xmark & \xmark \\
{\tt BAM:0017} & [96, 64]   & 0.985917 & \cmark & \xmark & \xmark & \xmark & \xmark & \xmark \\
{\tt BAM:0021} & [96, 64]   & 0.957098 & \xmark & \xmark & \xmark & \xmark & \xmark & \xmark \\
{\tt BAM:0037} & [216, 144] & 0.998790 & \cmark & \cmark & \cmark & \xmark & \xmark & \xmark \\
{\tt BAM:0048} & [108, 72]  & 0.983724 & \xmark & \xmark & \xmark & \xmark & \xmark & \xmark \\
{\tt BAM:0058} & [64, 64]   & 0.999127 & \cmark & \cmark & \cmark & \xmark & \xmark & \xmark \\
{\tt BAM:0064} & [240, 160] & 0.997427 & \cmark & \xmark & \cmark & \xmark & \xmark & \xmark \\
{\tt BAM:0091} & [144, 108] & 0.997810 & \cmark & \cmark & \cmark & \xmark & \xmark & \xmark \\
{\tt BAM:0094} & [144, 108] & 0.996804 & \cmark & \xmark & \cmark & \xmark & \xmark & \xmark \\
{\tt BAM:0095} & [256, 192] & 0.999550 & \cmark & \cmark & \cmark & \cmark & \cmark & \xmark \\
{\tt BAM:0107} & [128, 96]  & 0.995219 & \cmark & \xmark & \xmark & \xmark & \xmark & \xmark \\
{\tt BAM:0127} & [128, 96]  & 0.999011 & \cmark & \cmark & \cmark & \xmark & \xmark & \xmark \\
\hline\hline
\end{tabular}
\end{table}

Numerical Relativity (NR) simulations are fundamental for
understanding the the merger physics and the waveform morphology in
the high-frequencies regime.
They incorporate hydrodynamical effects, and can model not only the 
late-inspiral-merger parts of the coalescence, but also the postmerger
phase. 
While NR waveforms are often regarded as exact with respect to the 
ones provided by waveform approximants in the same regime, the complex
3D simulations can introduce significant uncertainties,
e.g. \cite{Bernuzzi:2011aq,Bernuzzi:2012ci,Radice:2013hxh,Radice:2016gym,Bernuzzi:2016pie}.  
The latter are both due to systematics (finite radius extraction of the GWs,
numerical dissipation, etc.) and to finite grid
resolution. Systematics are difficult to control, but finite
resolution errors can be studied by simulating at different
resolutions and performing convergence tests.

In this appendix, we apply the method of Sec.~\ref{sec:PEbiases}
to a set of NR waveforms taken from the {\tt CoRe} database \cite{Dietrich:2018phi}, 
with the aim of testing the accuracy of current state-of-the-art NR simulations
and guiding future effors. 
In particular, we consider multi-orbit and eccentricity reduced
simulations performed wih the BAM code, and focus on late
inspiral-merger where waveforms are shown to be convergent.
To the best of our knowledge, accuracy standard for BNS NR waveforms
at multiple grid resolutions have been computed only in \cite{Bernuzzi:2011aq} for data that are
currently superseded by the those produced with simulations employing
high-order numerical fluxes \cite{Radice:2013hxh,Bernuzzi:2016pie} and
higher resolutions that we consider here.  
We use here a sample of {\tt CoRe} waveforms computed at multiple
resolution and produced in \cite{Bernuzzi:2014kca, Dietrich:2016hky, Dietrich:2017aum, Dietrich:2017feu, Dietrich:2017xqb}.

Table~\ref{tab:nr_acc} displays the faithfulness values computed for a
set of BAM waveforms. Each value is obtained by comparing the two
highest-resolution simulations available  
for each considered set of intrinsic parameters. For each resolution $R$,
the simulations compute the multipoles $h_{\ell m}(t)$;
the waveform polarizations $h_{+}^R, h_{x}^R$ are reconstructed from
\be 
h_{+}-i h_{\times} =D_{L}^{-1} \sum_{\ell=2}^{\infty} \sum_{m=-\ell}^{\ell} h_{\ell m}(t)_{-2} Y_{\ell m}(\iota, \psi)
\ee
where$_{-2} Y_{\ell m}(\iota, \psi)$ are the spin weighted spherical
harmonics of spin $s=-2$ and 
$D_{L}$ is the luminosity distance. 
Assuming for simplicity that the radiation is emitted along the $z$-axis, 
perpendicular to the orbital plane, one has that $\iota = \psi = 0$,
and $_{-2} Y_{2\pm2}(0, 0) = \sqrt{5/(64 \pi)} (1 \pm 1)^2$. 
Fixing the source in GW170817's sky location and projecting the polarizations 
on the Livingston detector, matches are finally computed 
over a frequency range $f \in [f_{\rm low}, f_{\rm mrg}]$, where $f_{\rm low}$ is defined as
the frequency at which the amplitude of the fourier transform ${\rm FT}[{\rm Re}(h_{22})]$ is highest and $f_{\rm mrg}$ is the 
merger frequency, i.e the istantaneous frequency corresponding to the peak of the amplitude $|h_{22}(t)|$. Such values are then contrasted to the threshold faithfulness 
$\mathcal{F}_{\rm thr}$ of Eq.~\ref{eq:LB}. Similarly to Sec.~\ref{sec:},
we choose $\epsilon^2$ to be equal to one, for a stricter requirement, or to the number of intrinsic 
parameters of a BNS system $(N=6)$.
Note that while $\mathcal{F} < \mathcal{F}_{\rm thr}$ is a necessary but not sufficient condition for biases to appear, 
$\mathcal{F} >  \mathcal{F}_{\rm thr}$ is a sufficiently strong
requirement to ensure that two waveforms are faithful.
While for low SNR signals most of the waveforms considered are
accurate enough, we find that -- out of the twelve simulations
examined -- none passes the accuracy test when (SNR 80, $\epsilon = 1$), 
and only one ({\tt BAM:0095}) manages to pass it when
(SNR 80, $N=6$) and (SNR 30, $\epsilon=1$). Note the stars are
resolved in this case with ${\gtrsim}200$ grid points.

Our findings indicate that the largest portion of the NR
simulations available to date may not be yet sufficiently accurate for GW
data-analysis purposes. High-order methods for hydrodynamics and
resolutions ${>}200$ grid points per star appear necessary for GW modeling.

\section{Universal Relations}
\label{app:UR}

In this appendix we collect the quasi-universal relations employed in the 
main text.
\begin{itemize}

\item \textbf{De et. Al:}
This phenomenological relation was first introduced in \cite{De:2018uhw}, and links
the chirp mass of a BNS system $\mathcal{M}_c$ and its mass-weighted tidal
parameter $\tLam$ to the radius of a $1.4 \Msun$ star, $R_{1.4}$.
Explicitly,
\be 
\label{eq:De}
R_{1.4} \simeq(11.2 \pm 0.2) \frac{\mathcal{M}_c}{M_{\odot}}\left(\frac{\tLam}{800}\right)^{1 / 6} \mathrm{km}
\ee
Note that this expression is valid for GW170817-like systems, and is 
expected to fail for stars lighter than $1 \Msun$ or heavy systems, with chirp 
mass $\mathcal{M}_c$ larger than $1.5$.

\item \textbf{Binary Love and C-Love:}
These relations were obtained in \cite{Yagi:2015pkc} and \cite{Yagi:2016bkt}.
The Binary-Love relation links the asymmetric combination of the
tidal parameters $\Lambda_a = (\Lambda_1 - \Lambda_2)/2$ to the symmetric one 
$\Lambda_s = (\Lambda_1 + \Lambda_2)/2$ in a mass-ratio ($q$) dependent way:
\begin{align} 
  \label{eq:BinaryLove}
  \Lambda_a = & F_{n}(q) \Lambda_s \frac{a+\sum_{i=1}^{3} \sum_{j=1}^{2} b_{i j} q^{j} {\Lambda}_{s}^{-i / 5}}{a+\sum_{i=1}^{3} \sum_{j=1}^{2} c_{i j} q^{j} {\Lambda}_{s}^{-i / 5}} \\
  F_{n}(q) \equiv & \frac{1-q^{10 /(3-n)}}{1+q^{10 /(3-n)}}
\end{align}
where the coefficients $n, a, b_{i j}, c_{i j}$ can be found in e.g \cite{Yagi:2015pkc}.
It can be used in PE to reduce the dimensionality of the parameter space 
by linking $\Lambda_2$ and $\Lambda_1$ \cite{Chatziioannou:2018vzf}.
The C-Love relation, instead, links the compactness of a NS to its tidal deformability:
\be
C_i(\Lambda_i)= \sum_{k=0}^{2} a_{k}\left(\ln {\Lambda_i}\right)^{k}
\label{eq:CLambda}
\ee
and $a_0=0.3616998$, $a_1=-0.0354818$, $a_2=0.0006193849$.
To obtain an estimate of the radius of one of the NSs we combine them.
Indeed, rather than directly applying Eq.~\eqref{eq:CLambda} to the 
posterior samples of $\Lambda_2$, we wish to map $\tLam$ into $R$, as $\tLam$ is
the better measured quantity from GW analysis. To do so,
we obtain the relation $\Lambda_1 = \Lambda_1(\Lambda_2, q)$ from the 
inversion of Eq.~\eqref{eq:BinaryLove}, and compute $\Lambda_2 = \Lambda_2(\tLam, q)$ 
from the definition of $\tLam$. Finally, we apply Eq.~\eqref{eq:CLambda}.

\item \textbf{Raithel et al.}
The relation found in \cite{Raithel:2019uzi} is based on a quasi-newtonian approximation of the full 
relativistic expression for the tidal deformability of a stare, given by equation 
(96) of \cite{Damour:2009vw} with $\beta \approx 1$. Explicitly:
 \begin{align}
 \label{eq:Raithel}
      \tLam     = & \tLam_{0}\left(1+\delta_{0}(1-q)^{2}\right)\\
      \tLam_{0} = & \frac{15-\pi^{2}}{3 \pi^{2}} \xi^{-5}(1-2 \xi)^{5 / 2}\\
      \delta_{0}= & \frac{3}{104}(1-2 \xi)^{-2}\left(-10+94 \xi-83 \xi^{2}\right)\\
      \xi       = & \frac{2^{1 / 5} G \mathcal{M}_{c}}{R c^{2}} 
\end{align}
The above equations can be inverted numerically to obtain $R(\tLam, q, \mathcal{M}_c)$.
\end{itemize}

\section{$Q_\omega$ analysis with other approximants}
\label{app:SelfSpin}

In the present appendix we repeat the discussion of the second part of Sec.~\ref{sec:wf},
and compute $\Delta Q_{\hat\omega}$, $\Delta Q_{\hat\omega}^T$ and $\Delta Q_{\hat\omega}^{PM}$ 
for two additional state of the art approximants: (i) {\tt IMRPhenomPv2NRTidalv2}, which 
differs from the $\NRT1$ model exclusively in its tidal sector, 
which now incorporates a 7.5PN low frequency limit and PN-expanded spin-quadrupole
interactions up to 3.5PN in the waveform phase; 
and (ii) {\tt TaylorF2} endowed with quasi-5.5PN point mass terms, 7.5PN tides
and spin-spin terms up to 3.5 PN. 
We note it is not possible to obtain meaningful information from $Q_\omega$ for
{\tt SEOBNRv4Tsurrogate}, as it is not C1 over the whole 
range of frequencies considered.

\begin{figure*}
  \centering 
  \includegraphics[width=0.8\textwidth]{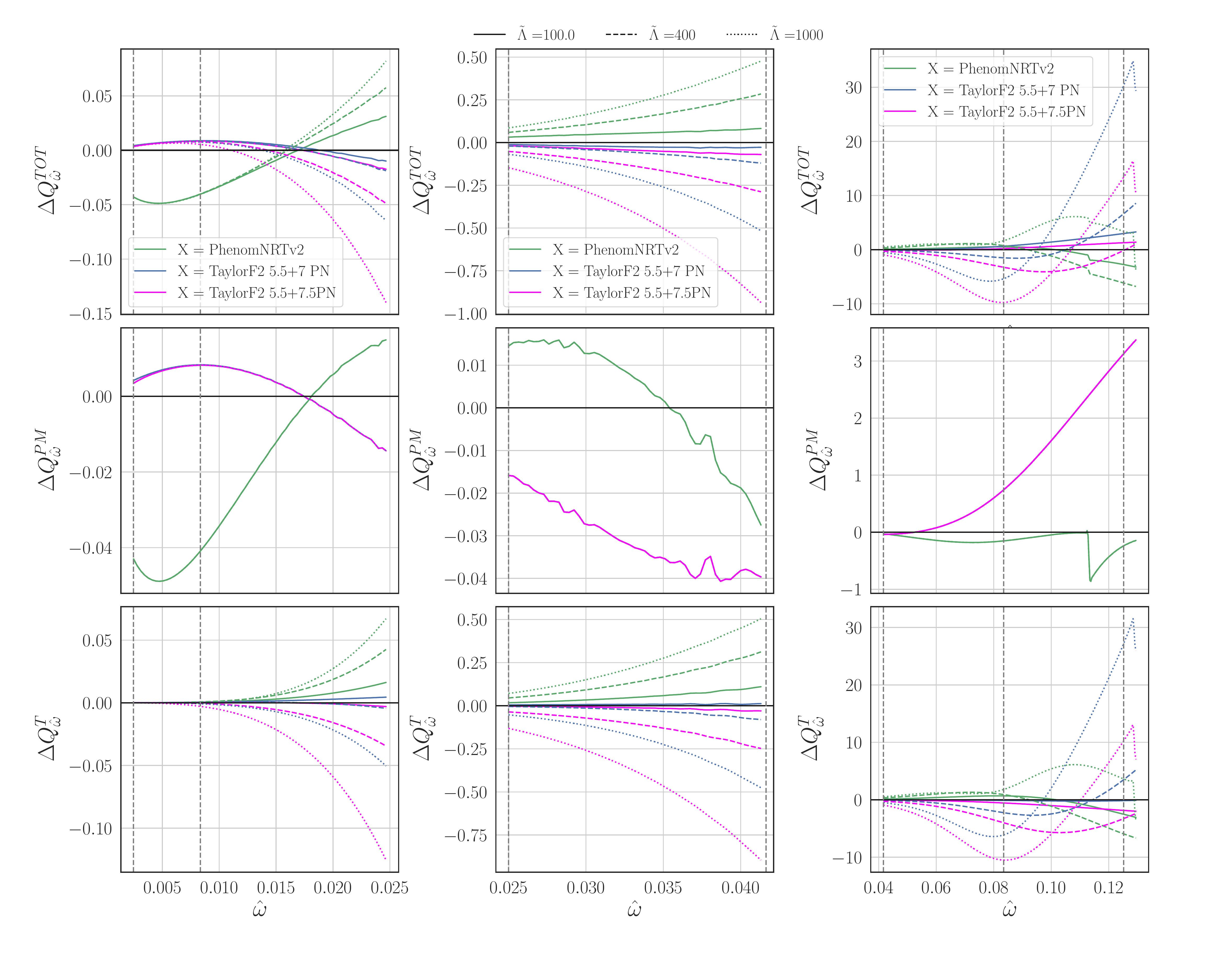}
    \caption{This Figure is the equivalent of Fig.~\ref{fig:Qomg} with {\tt TaylorF2}
     at 5.5PN PM and 7 PN tides (blue), 5.5PN PM and 7.5PN tides (magenta) and {\tt IMRPhenomPv2NRTidalv2} (green).
      Note that the 5.5PN PM and {\tt NRTidalv2} descriptions improve the $\Delta Q_{\hat{\omega}}^{PM}$ and $\Delta Q_{\hat{\omega}}^{T}$ in the low and high frequency regimes, respectively. When coupled to a 5.5PN PM, 7.5 PN tides are closer to {\tt TEOBResumS} for $\tLam = 100$ than the 7PN description, which in turn performs better for $\tLam = 400$ over a large portion of the frequency range.}
    \label{fig:Qomg_app}
\end{figure*}

Figure~\ref{fig:Qomg_app} shows $Q_{\hat\omega}$ for the aforementioned approximants, computed
for three reference signals with varying $\tLam$ and zero spins, once again divided into 
three $\hat{\omega}$ intervals, correspoding to the regimes in which tidal contributions
are roughly smaller than, comparable to or dominant with respect to $\Delta Q_{\hat\omega}^{PM}$.
Inspecting the first column (which corresponds to the early frequencies interval),
we notice that the $\Delta Q_{\hat\omega}^{PM}$ of {\tt TaylorF2} is comparable 
to that of {\tt IMRPhenomPv2NRTidalv2}, and overall closer to {\tt TEOBResumS}' description than the 
one provided by considering a 3.5PN point mass baseline. When considering the
tidal sector, instead, we note that the behaviours of {\tt NRTidalv2} and 7.5PN tides
are opposites from the start.
Moving to higher frequencies, tidal effects dominate both the late inspiral and merger regimes
for both approximants. {\tt IMRPhenomPv2NRTidalv2} is less attractive than $\NRT1$, while 
the 7.5 PN tidal description is more repulsive than the 6PN one.
Close to merger we find that -- as expected -- the point mass contribution of the 5.5PN
approximant becomes large and positive, and partially compensates the negative $\Delta Q_{\hat\omega}^T$.
Overall, we find that {\tt IMRPhenomPv2NRTidalv2} provides a description that is much closer to {\tt TEOBResumS}'
than the one offered by $\NRT1$, albeit being still slightly more attractive.

\begin{figure*}
  \centering 
  \includegraphics[width=0.8\textwidth]{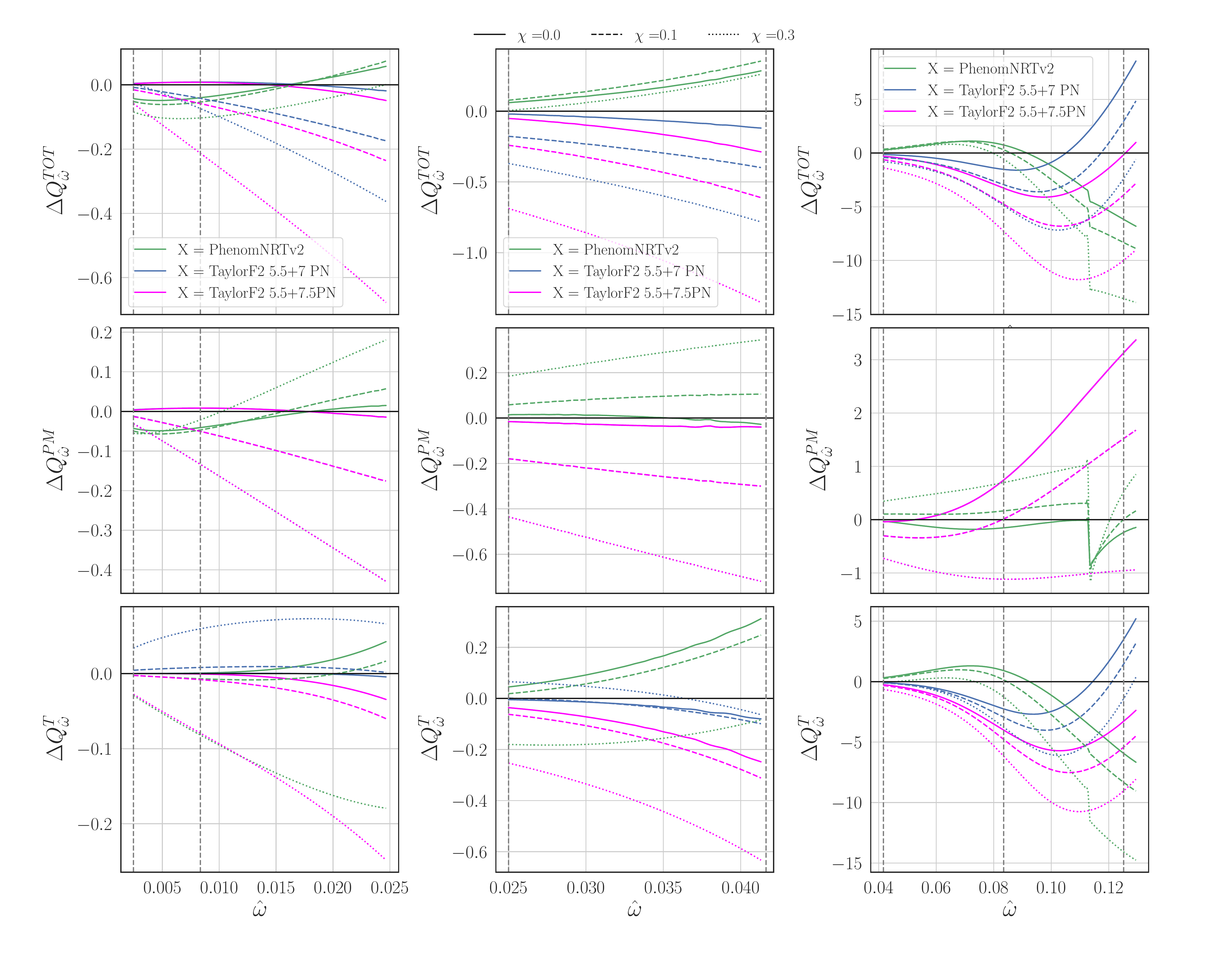}
    \caption{This Figure is the equivalent of Fig.~\ref{fig:Qomg} with {\tt TaylorF2}
     at 5.5PN PM and 7 PN tides (blue), 5.5PN PM and 7.5PN tides (magenta) and {\tt IMRPhenomPv2NRTidalv2} (green).
     We note that in the low frequency regime the 5.5PN PM, which gives the best approximation of the $\TEOB$ PM
     between the approximants considered for nonspinning equal-mass binaries, becomes increasingly more negative as 
     spins grow. On the other hand, the Phenom description over the same range consistently has 
     $|\Delta Q_{\hat{\omega}}^{PM}| < 0.2$. When considering the high frequency contributions, instead, the hierarchy 
     displayed in Fig.~\ref{fig:Qomg_spin} is mantained.}     
  \label{fig:Qomg_app_spin}
\end{figure*}

Spinning configurations are studied in Fig.~\ref{fig:Qomg_app_spin}, which 
shows $\Delta Q_{\hat\omega}$ for three target signals with $\tLam = 400$ and 
increasing spins $\chi_1=\chi_2$. 
Focusing on the low frequency contribution to $\Delta Q_{\hat{\omega}}$, we observe that
{\tt IMRPhenomPv2NRTidalv2} is now overall closer to {\tt TEOBResumS} than {\tt IMRPhenomPv2NRtidal}
and {\tt TaylorF2} were. Additionally, the 5.5 PN point mass, which in the non-spinning case follows closely the
behaviour of {\tt TEOBResumS}, becomes increasingly more negative as spins grow.
Moving to the high frequency regime, the improvements of {\tt NRTidalv2} have a positive
effect on $\Delta Q_{\hat{\omega}}^T$ of the phenomenological approximant. Indeed, while the Phenom-EOB-PN
hierarchy displayed in Fig.~\ref{fig:Qomg_spin} for spinning binaries is mantained, the differences decrease and {\tt NRTidalv2} is closer to {\tt TEOBResumS}' description than {\tt NRTidal}.

\label{sec:}

\end{document}